\documentclass{article}
\usepackage{graphicx} % Required for inserting images
\usepackage[utf8]{inputenc}
\usepackage{amsmath,amssymb}
\usepackage{stmaryrd}
\usepackage{bbm}

\DeclareMathOperator*{\argmax}{arg\,max}
\DeclareMathOperator*{\argmin}{arg\,min}
\usepackage[ruled]{algorithm2e}
\usepackage{hyperref}
\usepackage{cleveref}

\usepackage{amsthm}
\newtheorem{lemma}{Lemma}

\newtheorem{definition}{Definition}
\newtheorem{corollary}{Corollary}

\newtheorem{theorem}{Theorem}

\newtheorem{remark}{Remark}
\usepackage{fullpage}

\usepackage{graphicx}
\usepackage{subcaption}

\usepackage{comment}
\usepackage{mathtools}

\usepackage[shortlabels]{enumitem}

\usepackage[shortlabels]{enumitem}

\newcommand{\ignore}[1]{}

\newcommand{\hide}[1]{}

\usepackage[shortlabels]{enumitem}

\newcommand{\coloneqq}{:=}

\usepackage[shortlabels]{enumitem}

\newcommand{\A}{\mathcal{A}}

\usepackage{caption}
\usepackage{xcolor}

\DeclareMathOperator{\sgn}{sgn}

\newcommand{\cA}{\mathcal{A}}

\newcommand{\cF}{\mathcal{F}}
\newcommand{\cG}{\mathcal{G}}

\newcommand{\cM}{\mathcal{M}}

\newcommand{\cP}{\mathcal{P}}

\DeclareMathOperator{\BR}{\mathsf{BR}}

\newcommand{\reals}{\mathbb{R}}

\newcommand{\co}{\overline{\operatorname{co}}}
\newcommand{\ext}{\operatorname{ext}}
\newcommand{\cl}{\operatorname{cl}}
\newcommand{\upop}{[\underline{p}, \overline{p}]}

\usepackage{xspace}
\newcommand*{\eg}{e.g.\@\xspace}
\newcommand*{\ie}{i.e.\@\xspace}

\newcommand*{\cf}{cf.\@\xspace}
\usepackage{natbib}

\title{The Value of Ambiguous Commitments in Multi-Follower Games}

\author{Natalie Collina \and Rabanus Derr \and Aaron Roth}
\date{\today}

\begin{document}

\maketitle

\begin{abstract}
We study games in which a leader makes a single commitment, and then multiple followers (each with a different utility function) respond. In particular, we study \emph{ambiguous commitment strategies} in these games, in which the leader may commit to a \emph{set} of mixed strategies, and ambiguity-averse followers respond to maximize their worst-case utility over the set of leader strategies. Special cases of this setting have previously been studied when there is a single follower: in these cases, it is known that the leader can increase her utility by making an ambiguous commitment if the follower is restricted to playing a pure strategy, but that no gain can be had from ambiguity if the follower may mix. We confirm that this result continues to hold in the setting of general Stackelberg games.

We then develop a theory of ambiguous commitment in games with multiple followers. We begin by considering the case where the leader must make the same commitment against each follower. We establish that --- unlike the case of a single follower --- ambiguous commitment can improve the leader's utility by an unboundedly large factor, even when followers are permitted to respond with mixed strategies. This result holds even in simple zero-sum games. We go on to show an advantage for the leader coupling the same commitment across all followers, \emph{even when she has the ability to make a separate commitment to each follower}. In particular, there exist general sum games in which the leader can enjoy an unboundedly large advantage by coupling her ambiguous commitment across multiple followers rather than committing against each individually. In zero-sum games we show there can be no such coupling advantage. Finally, we give a polynomial time algorithm for computing the optimal leader commitment strategy in the special case in which the leader has $2$ actions (and $k$ followers may have $m$ actions), and prove that in the general case, the problem is NP-hard.
\end{abstract}

 \thispagestyle{empty} \setcounter{page}{0}
 \clearpage

 \thispagestyle{empty} \setcounter{page}{0}
 \clearpage

\section{Introduction}

In the standard Stackelberg game model, a leader commits to a distribution over her actions in order to maximize her utility against a best-responding follower. In this work, we propose adding a new tool to the leader's arsenal: deliberate ambiguity. Instead of committing to a single distribution, a leader may commit to an \emph{ambiguous set} of distributions. This ambiguity may be implicit in the commitment itself (such as ambiguous wording in a law or a contract), or may be explicit in the design of the commitment (such as defining a set of leader actions and delegating the choice from this set to a third party with unknown preferences). Behavioral studies have shown that, when faced with ambiguity, humans act in order to minimize their worst-case outcome~\cite{pulford2007ambiguous}. We study the performance of ambiguity against followers who are \emph{ambiguity-averse} in this sense.

There has been some recent work studying ambiguous commitments against ambiguity-averse followers in the special case of principal-agent contracts with a single agent \cite{dütting2024ambiguouscontracts} and buyer-seller models \cite{tillio2016design}. These works show that ambiguity can be strictly beneficial for the leader if the follower is obligated to respond with a pure strategy --- and that the benefit disappears if the follower is allowed to play mixed strategies. While responding with pure strategies is without loss of generality in classical models, in the case of an ambiguity-averse follower, his true optimal response will often be a mixed strategy. 

In this work, we study the general case of Stackelberg games (which generalizes previously studied principal-agent contracting models and buyer-seller games), and consider the value of ambiguity when there can be multiple followers. We first show that the results of \cite{dütting2024ambiguouscontracts,tillio2016design}, showing that ambiguity is not a useful tool against a single follower equipped with mixed strategies, generalize to any Stackelberg game. However, we go on to show that there remain focal leader-follower settings in which ambiguity is a beneficial commitment tool. We study a \emph{coupled Stackelberg} setting, in which a leader must select a single strategy to commit to against $k$ ambiguity-averse followers, each playing a separate $n$ by $m$ game with the leader. This setting is motivated by the fact that commitments are commonly seen by many independent downstream agents---for example, a randomized security policy in a Stackelberg security game will not be seen by only one bad actor, but multiple bad actors with different preferences and different abilities to cause harm. In this setting, we show that committing to an ambiguous strategy can be strictly beneficial, even if the ambiguity-averse agents can use mixed strategies. 

Furthermore, we show that, perhaps surprisingly, there are multi-follower Stackelberg games in which committing to the same ambiguous strategy against all followers is better even than committing to the optimal Stackelberg distribution in each game \emph{separately}. This implies that ambiguous commitments are useful not only in coupled Stackelberg games, but also in uncoupled settings in which the leader has commitment power in multiple games simultaneously. Instead of solving these games separately, it may be strictly beneficial for them to commit to the same ambiguous strategy in all of them.\footnote{If the leader has a different number of actions in each game, she can simply duplicate actions in order to have a consistent dimension. The number of actions of each follower does not need to be the same.} 

This may seem counterintuitive; we show in this paper that ambiguity has no benefit in decoupled Stackelberg games (Theorem~\ref{thm:single follower - no ambiguity needed}), and that coupling has no benefit in Stackelberg games with classic commitment (Lemma~\ref{lem:ISV_lb}). As unilaterally coupling the game or adding the power of ambiguity never helps the leader, why should it help to do both in tandem?

For intuition, consider a setting where a leader has a set of costly ``verification" actions, each of which discourage different types of downstream agents from behaving poorly. It turns out that the threat of this verification can be much more efficiently utilized in a coupled, ambiguous setting. Take the example of an airport security authority which must design and communicate their daily security measures with regard to $k$ different types of security threats. First, consider the decoupled setting in which the authority communicates a separate policy for each threat type. In order to discourage all threats, the authority must, with some reasonable probability, exert effort each day to verify that the airport is safe from every threat $i$. In a coupled, ambiguous Stackelberg game, by contrast, the authority can include the action of spot-checking against every threat type in their ambiguous set. All ambiguity-averse agents who pose a threat to the airport will be discouraged regardless of their threat type--but as the authority's action is now coupled across all threat groups, only one verification action will be taken per day, saving effort for the authority.
%\rab{really nice, now! Like it. thought about some halfsentence which says something like: "The authority hedges against the ambiguity by coupling." (the agent's in fact cant)}

Finally, we study the computational complexity of ambiguous commitment. We give an efficient algorithm for finding the optimal ambiguous strategy to commit to in coupled Stackelberg games when the number of leader actions is $n = 2$ (and both the number of followers $k$ and the number of follower actions $m$ may grow). This captures many central leader-follower models such as linear contracts (which have been shown to be robust to ambiguity in many settings) and $1$-dimensional Bayesian persuasion.\footnote{This is because $1$-dimensional \emph{linear} utility functions can be thought of as taking expectations over just two actions.} We complement these results by showing that finding the optimal ambiguous commitment strategy is NP-hard in the general case when the number of leader actions may be as large as the number of followers: $n = \Omega(k)$.

\subsection{Overview of Our Contributions} 

\begin{itemize}
\item In Section~\ref{sec:model}, we formally introduce the notion of a coupled Stackelberg game, in which a single leader makes the same commitment (which may or may not be ambiguous) against multiple downstream followers. 
\item In Section~\ref{sec:ambiguity}, we characterize the benefits of ambiguity in coupled Stackelberg games. We define a notion of an \emph{ambiguity gap} of a coupled game, which is the ratio between the best payoff obtainable by a leader who may make ambiguous commitments and the best payoff of a leader who must commit to a single (non-ambiguous) mixed strategy in a coupled Stackelberg game. If the number of followers in the game is $1$ (which reduces the game to a standard 2-player Stackelberg game), we show that the ambiguity gap is $1$; in other words, there is no benefit of ambiguity against a single follower in any Stackelberg game (Theorem~\ref{thm:single follower - no ambiguity needed}). On the other hand, if there are even $2$ followers, there are games where the ambiguity gap is unbounded, even in the special case of zero-sum games (Theorem~\ref{thm:amb_gap_zs}). These results are summarized in the left column of Table~\ref{tab:benefit_amb}. 
\item In Section~\ref{sec:coupling}, we consider a leader playing multiple Stackelberg games who need not make the same commitment in all games, but could choose to if she so desired.\footnote{This is possible in all settings where the commitments can be made simultaneously.} We define a new notion called the \emph{ambiguous coupling gap}, which is the ratio between the utility of the best ambiguous commitment if the leader chooses to play a correlated commitment across all games, and the sum of utilities of the best individual commitments against each follower separately. We show that, when all the games are zero-sum, there is no benefit to coupling and playing ambiguously (Theorem~\ref{thm:Bounds to Ambiguous Stackelberg Value}). We go on to show that, when games are allowed to be general-sum, there are simple examples of games (2 follower and 2 actions) where the ambiguous coupling gap is unbounded; in other words, it is better to artificially couple the games and play an ambiguous commitment set (Theorem~\ref{thm:coupling_gap}). These results are summarized in the right column of Table~\ref{tab:benefit_amb}.  
\item In Appendix~\ref{app:structure}, we motivate the difficulty of the problem of computing an optimal ambiguous commitment by showing that the optimal ambiguous commitment need not be the set of all pure strategies (Lemma~\ref{thm:pure_unbounded}) and in fact may have extreme points which are mixed strategies (Lemma~\ref{thm:frac_opt}). Notably, the optimal commitment set can include mixed strategies even when the number of leader actions $n=2$.  
\item In Section~\ref{sec:poly-time}, show that, despite this complexity, there is a wide class of games for which we can compute good ambiguous commitments. In Theorem~\ref{thm:poly-time algorithm for 2x m games}, we give an algorithm for computing an approximately optimal ambiguous commitment in multi-follower games where $n=2$, which runs in time polynomial in the number of follower actions $m$ and number of followers $k$.
\item In Section~\ref{sec:hardness}, we show that finding the optimal commitment in the general case (where it is possible for $n = \Omega(k)$) is NP-hard via a reduction to min-vertex cover. 
\end{itemize}

 \begin{table}[h]
        \centering
        \begin{tabular}{c|c c}
             & Max Ambiguity Gap & Max Ambiguous Coupling Gap \\ \hline
            Single-follower & 1 (Theorem~\ref{thm:single follower - no ambiguity needed}) &  1 (Theorem~\ref{thm:single follower - no ambiguity needed})\\
            Multi-follower, zero-sum & $\infty$ (Theorem~\ref{thm:amb_gap_zs}) & 1 (Theorem~\ref{thm:Bounds to Ambiguous Stackelberg Value})\\
            Multi-follower, general-sum & $\infty$ (Theorem~\ref{thm:amb_gap_zs}) & $\infty$ (Theorem~\ref{thm:coupling_gap}) 
        \end{tabular}
        \caption{The maximum ambiguity gap and ambiguous coupling gap of any game in the respective category.}
        \label{tab:benefit_amb}
    \end{table}

\section{Related Work}
The majority of research in game theory and decision theory  revolves around the expected utility framework--the assumption that rational agents will act to maximize their expected utility under a known distribution of states. However, this framework fails to provide a rationale for decision making under \emph{ambiguity}\footnote{Note that we use the term ``ambiguity'' similar to Knight's use of ``uncertainty'' \citep{knight1921risk}.} \citep{ellsberg1961risk}. This led to the maxmin-expected utility framework introduced in \citet{gilboa1989maxmin}. These insights and technical extensions soon were adopted in game-theoretic settings \citep[Section 4]{mukerji2004overview}.

The strategic impact of ambiguity has been explored in simultaneous-action games when agents are endowed with ambiguous beliefs about their payoffs or about the other agents' strategies \citep{perchet2014note}. Another line of work, more related to our own, has equipped agents in simultaneous games with ``ambiguization'' devices which extended their action space from a single mixed strategy to a set of mixed strategies \citep{bade2011ambiguous, riedel2011strategic, riedel2014ellsberg}.\footnote{There is an even earlier note towards this idea \citep{binmore2007rational}, in which ambiguous strategies are called ``multiplex strategies''.} For a nice summary of games with simultaneous actions and ambiguity, see Linda Sass' PhD thesis \citep{sass2013ellsberg}.

An analogous development can be observed in sequential commitment games. Relatively recently, \citep{liu2018distributionally, kroer2018robust} and \citep{gan2023robust} discussed Stackelberg game setups in which leader's knowledge about follower's payoffs or strategies is ambiguous. By contrast, in our work the leader has full knowledge of the game, and may (or may not) choose to utilize ambiguity as a tool.

More similarly to our work, \citet{tillio2016design, beauchene2019ambiguous} and \citet{dütting2024ambiguouscontracts} formalize specific sequential commitment games in which a leader is allowed to commit to a set of strategies. Their work focuses on mechanism design, Bayesian persuasion, and contracting, respectively.

In this work, we turn to general Stackelberg games introducing ambiguous leader commitments here. Our work puts the focus differently in several aspects: (a) We assume that the followers respond with potentially mixed strategies to ambiguous commitments by a leader. (b) We assume that the leader optimizes her expected worst-case payoff over its ambiguous commitment set. (c) Our investigation mainly considers coupled and decoupled multi-follower setups. A multi-follower Stackelberg game is coupled if a single leader's commitment is broadcast to multiple followers. It is decoupled if a single leader commits to different strategies for each follower individually.

\citet{beauchene2019ambiguous} share elements (a) and (b). However, the particular payoff structure of ambiguous persuasion makes their setting incomparable. They concentrate on single-follower persuasion. \citep{tillio2016design} and \citep{dütting2024ambiguouscontracts} also well focus on single-follower settings. In particular, they assume, differently for us, that the follower responds with pure actions. This assumptions turns out to be crucial for not trivializing the single-follower setup. As proven in \citep{dütting2024ambiguouscontracts} for contracting games and as we show for general Stackelberg games, if a single agent responds via a mixed strategy to a leader's ambiguous commitment, then there is no advantage in employing ambiguous commitments for the leader. This sparked our interest in multi-follower settings.

Finally, \citep{tillio2016design} and \citep{dütting2024ambiguouscontracts} require the ambiguous commitment sets to be ``consistent''. Consistency requires that for each best response the leader's expected payoff is equal for each strategy in the ambiguous commitment. Instead of placing this restriction on the leader's commitment sets, we consider the leader's utility in the worst case over her commitment set. %Surprisingly, \citep{beauchene2019ambiguous} show that for ambiguous persuasion in their framework it is necessary and sufficient to use ``weak synonyms'', which imply the ambiguous commitment set to be consistent.\ar{Do we need this last sentence? It doesn't add much for a reader like me who doesn't know what a weak synonym is}\rab{not necessarily. havent seen this statement around though. was just part of trying to embed the works in a web.}

The computation of Stackelberg strategies in a simple full-information, two-follower setup has been solved by \citet{conitzer2006computing}, who show an efficient algorithm for computing the leader's optimal commitment distribution. To the best of our knowledge, this is the first study of ambiguous commitments in general Stackelberg games. %In particular, we extend previous work to multi-follower setups, raising new and interesting questions.

Our fundamental behavioral assumption that the involved agents are ambiguity-averse is backed-up conceptually \citep{ellsberg1961risk}, as well as experimentally \citep{pulford2007ambiguous}.

\section{Model}
\label{sec:model}
\subsection{Coupled Stackelberg Game}

 We consider a sequential game with one leader $L$ and several followers $F \in \cF$ ($|\mathcal{F}| < \infty$), each having finite action spaces $A_L$ and $(A_F)_{F \in \cF}$. We use $n$ to denote the cardinality of the leader's action space and $m$ to denote the cardinality of the follower's action spaces (which without loss of generality via action duplication can be taken to be of equal size).  For every follower there is a utility function $u_F \colon A_L \times A_F \rightarrow \reals$ for the follower and the leader $u_{L_F} \colon A_L \times A_F \rightarrow \reals$. The leader's overall utility is defined by summing over her utility functions across all followers: $u_L(a_L, (a_F)_{F \in \mathcal{F}}) \coloneqq \sum_{F \in \mathcal{F}} u_{L_F}(a_L, a_F)$.

    For mixed strategies, we introduce the following notation for the expected payoff:
    \begin{align*}
        U_F(p_L, p_F) \coloneqq \mathbb{E}_{a_L\sim p_L, a_F \sim p_F}[u_F(a_L, a_F)]
    \end{align*}
    as well as
    \begin{align*}
        U_{L_F}(p_L, p_F) \coloneqq \mathbb{E}_{a_L\sim p_L, a_F \sim p_F}[u_{L_F}(a_L, a_F)].
    \end{align*}
    With slight abuse of notation, we write $a_L$ (respectively $a_F$) for the Dirac-distribution on $a_L$ (respectively $a_L$).

    \begin{definition}[Coupled Stackelberg Game]
    \label{def:coupled Stackelberg Game}
        We call $\cG = (\cF, A_L, (A_F)_{F \in \cF}, (u_{L_F})_{F \in \cF}, (u_F)_{F\in \cF})$ a \emph{coupled Stackelberg game} if the leader commits to a strategy which is broadcast to all followers. Each follower $F \in \cF$ responds to that strategy with an element from a set of responses.
        If $u_{L_F} = -u_F$ for all $F \in \cF$ then we call $\cG$ a \emph{coupled, zero-sum Stackelberg game}. If $|\cF| = 1$, then we simply call $\cG$ a \emph{Stackelberg game}.
    \end{definition}
    A coupled Stackelberg game is different from a decoupled Stackelberg game in which the leader commits to different strategies for each follower. Such a decoupled setting can be formalized by a tuple of coupled Stackelberg games with a single follower, \ie $(\cG_F)_{F \in \cF} = \left((\{F\}, A_L, A_F, u_{L_F}, u_F))\right)_{F \in \cF}$.
    We intentionally have left the type of commitment and the type of response undefined in this definition, and will discuss how to instantiate this setting with and without ambiguity below.

    In the standard definition of a Stackelberg game, the leader commits to a mixed strategy and each follower will best respond to it---i.e. chooses the action that maximizes his expected utility given the leader's commitment.
    \begin{definition}[Classical Commitment]
        We call $\cG = (\cF, A_L, (A_F)_{F \in \cF}, (u_{L_F})_{F \in \cF}, (u_F)_{F\in \cF})$ a \emph{coupled Stackelberg game with classical commitment} if the leader commits to a mixed strategy, \ie $p_L \in \Delta(A_L)$, where $\Delta(A_L)$ denotes the set of probability distributions on $A_L$. This strategy is broadcast to all followers. Each follower $F \in \cF$ responds to that mixed strategy with an element of the set of expected utility maximizing actions,\footnote{If the follower $F$ is clear from context we sometimes neglect writing $F$ out explicitly.} 
    \begin{align*}
        \BR_F(p_L) \coloneqq \argmax_{ p_F \in \Delta(A_F)} U_F(p_L, p_F) \subseteq \Delta(A_F).
    \end{align*}
    \end{definition}

    In order to define Stackelberg payoffs for the leader and follower, we have to introduce tie-breaking rules which select a single best response from the set of best responses.
    \begin{definition}[Tie-Breaking Rules]
        We call a general function $s \colon 2^{\Delta(A_F)} \rightarrow \Delta(A_F)$\footnote{We denote the power set of a set $S$ as $2^S$.}  such that $s(B) \in B$ for all $B \subseteq \Delta(A_F)$ a \emph{tie-breaking rule}.
        A tie-breaking rule is \emph{simple} if it is deterministic and selects only extreme points of the set of responses.\footnote{Hence, a simple tie-breaking rule is only well-defined if the set of responses has extreme points, which is the case if the set is compact \citep[Proposition 2.2.3]{hiriart2004fundamentals}. This is guaranteed by Lemma~\ref{lemma:Maxmin Responses are Closed and Convex}.} We call the tie-breaking rule $\Bar{s}$ which selects the most favorable element for the leader from the set of maximizing actions the \emph{strong tie-breaking rule}.\footnote{In combination with classical commitment this tie-breaking rule leads to the notion of a ``strong Stackelberg equilibrium'' \citep{loridan1996weak}.}
    \end{definition}
    For simplicity, we assume that every follower uses the same tie-breaking rule. In fact, most of our statements, except for the results in Section~\ref{sec:hardness}, are invariant to the choice of tie-breaking rule.

    For a given commitment $p_L \in \Delta(A_L)$ of the leader and any tie-breaking rule $s$, the payoff of follower $F \in \cF$ is 
    \begin{align*}
        \max_{a_F \in A_F} \mathbb{E}_{a_L \sim p_L}[u_F(a_L,a_F)] = U_F(p_L, s(\BR(p_L))).
    \end{align*}
    The payoff of the leader in a coupled Stackelberg game $\cG$ with tie-breaking rule $s$ and commitment $p_L \in \Delta(A_L)$ is
    \begin{align}
    \label{eq:leader payoff classical stackelberg game with multiple followers}
        V(s, p_L) &\coloneqq \mathbb{E}_{a_L \sim p_L}[ \mathbb{E}_{a_F \sim s(\BR_F(p_L)), F \in \cF}[u_L(a_L, (a_F)_{F \in \mathcal{F}}) ]] \\
        &= \mathbb{E}_{a_L \sim p_L}\left[ \mathbb{E}_{a_F \sim s(\BR_F(p_L)), F \in \cF}\left[\sum_{F \in \mathcal{F}}u_{L_F}(a_L, a_F) \right]\right]\\
        &= \sum_{F \in \mathcal{F}} U_{L_F}(p_L, s(\BR_F(p_L))).
    \end{align}
    
    \begin{definition}[Optimal Stackelberg Value and Stackelberg Equilibrium] 
        For a coupled Stackelberg game $\cG$ with classical commitment and a tie-breaking rule $s$ for the followers, we define the \emph{optimal Stackelberg value} as
        \begin{align}
            \label{eq:Stackelberg equilibrium}
            V^*(s) \coloneqq \max_{p_L \in \Delta(A_L)} V(s, p_L).
        \end{align}
        We call $(p_L^*, (\Bar{s}(\BR_F(p_L^*))_{F \in \cF})$ for
        \begin{align*}
            p_L^* \in \argmax_{p_L \in \Delta(A_L)} V(s, p_L),
        \end{align*}
        a \emph{Stackelberg equilibrium}.
    \end{definition}

\subsection{Stackelberg Games with Ambiguous Commitments}

So far, we have discussed the coupled Stackelberg setting without ambiguous commitment--this looks very similar to a standard Stackelberg game, simply with multiple followers engaging with the same commitment distribution. Now, we will extend this setting by allowing the leader to commit not just to a distribution, but to a set of distributions, \ie $P_L \subseteq \Delta(A_L)$, such that the way the true distribution will be drawn from the set is ambiguous. To do this, we must first clarify how the response of the follower should be defined. Second, we must specify how payoffs for both agents are calculated.

\begin{definition}[Ambiguous Leader Strategy]
    A leader's strategy is called \emph{ambiguous} if the leader commits to a non-empty, closed, convex set $P_L \subseteq \Delta(A_L)$. If $P_L$ is not a singleton, then the strategy is called \emph{strictly ambiguous}. We denote the set of all ambiguous strategies on $A_L$ as $\cA(A_L)$.
\end{definition}
We can restrict the commitments to closed, convex sets without loss of generality, as we later argue in Lemma~\ref{lemma:closed, convex Commitment Sets are Exhaustive}. We are now ready to instantiate our coupled game with ambiguous commitments. 

\begin{definition}[Ambiguous Commitment]
    We call $\cG = (\cF, A_L, (A_F)_{F \in \cF}, (u_{L_F})_{F \in \cF}, (u_F)_{F\in \cF})$ a \emph{coupled Stackelberg game with ambiguous commitment} if the leader commits to an ambiguous strategy $P_L \in \cA(A_L)$. This strategy is broadcast to all followers. Each follower $F \in \cF$ responds with an element of the set of maxmin responses,\footnote{If the follower $F$ is clear from context we sometimes neglect writing $F$ out explicitly.}
\begin{align*}
    \BR^a_F(P_L) \coloneqq \argmax_{ p_F \in \Delta(A_F)} \min_{p_L \in P_L} U_F(p_L, p_F).
\end{align*}
\end{definition}
The theory of maxmin responses to ambiguity dates back to the seminal work of \citep{gilboa1989maxmin}. Note that the maximum and minimum here are well-defined, as $\Delta(A_F)$ and $P_L$ are compact.

For a given ambiguous commitment $P_L \in \cA(A_L)$ of the leader and a tie-breaking rule $s$, the worst-case expected payoff for follower $F \in \cF$ is given by
\begin{align*}
    \max_{p_F \in \Delta(A_F)} \min_{p_L \in P_L} U_F(p_L, p_F) = \min_{p_L \in P_L} U_F(p_L, s(\BR^a(P_L)))
\end{align*}

Our next task is to define the payoff structure of the leader. While this is straightforward in classical commitment settings, where the realized action of the learner is clear, it is no longer clear here what outcome the leader should expect.\footnote{\citep{tillio2016design} and \citep{dütting2024ambiguouscontracts} address this issue by requiring that all of the commitments in the leader's ambiguous set give them the same utility against the maxmin best response of the follower. For a more detailed discussion see Section~\ref{sec:implementing ambiguity}.} We circumvent this issue by considering the leader's utility in the worst-case over her ambiguous set. This modeling choice ensures that any benefit of ambiguity we show is not the result of the leader gaming the system by committing to an ambiguous set and then picking her favorite action. Furthermore, it is exactly the utility function that an ambiguity-averse leader would aim to maximize if she delegated the selection from the set to a third party (\cf Section~\ref{sec:implementing ambiguity}).

The worst-case payoff of the leader in a coupled Stackelberg game $\cG$ with tie-breaking rule $s$ and commitment $P_L \in \cA(A_L)$ is
\begin{align}
\label{eq:leaders payoff}
    W(s, P_L) &\coloneqq \min_{p_L \in P_L} \mathbb{E}_{a_L \sim p_L}[ \mathbb{E}_{a_F \sim s(\BR^a_F(P_L)), F \in \cF}[u_L(a_L, (a_F)_{F \in \mathcal{F}}) ]]\\
&= \min_{p_L \in P_L} \mathbb{E}_{a_L \sim p_L}\left[ \mathbb{E}_{a_F \sim s(\BR^a_F(P_L)), F \in \cF}\left[\sum_{F \in \mathcal{F}}u_{L_F}(a_L, a_F) \right]\right]\\
    &= \min_{p_L \in P_L} \sum_{F \in \mathcal{F}}  U_{L_F}(p_L, s(\BR^a_F(P_L))).
\end{align}
Note that if we can provide strategies for the leader such that this pessimistic payoff is maximized, we directly provide a lower bound to the expected payoff for a less ambiguity-averse leader.

\begin{lemma}[Maxmin Responses are Closed and Convex]
\label{lemma:Maxmin Responses are Closed and Convex}
Let $P_L \in \cA(A_L)$. The set of maxmin follower responses 
  $ \BR^a_F(P_L) \coloneqq \argmax_{ p_F \in \Delta(A_F)} \min_{p_L \in P_L} U_F(p_L, p_F)$  
  is closed and convex.
\end{lemma}
\begin{proof}
First, we show that the set $\BR^a_F(P_L)$ is closed. Note, that for any sequence of $(p_F^i)_{i \in \mathbb{N}}$ such that $p_F^i \in \BR^a_F(P_L)$ for all $i \in \mathbb{N}$ and $\lim_{i \rightarrow \infty} p_F^i = \bar{p}_F \in \Delta(A_F)$, it holds
\begin{align*}
    \lim_{i \rightarrow \infty} \min_{p_L \in P_L} U_F(p_L, p^i_{F}) = \min_{p_L \in P_L} \lim_{i \rightarrow \infty} U_F(p_L, p^i_{F}) = \min_{p_L \in P_L} U_F(p_L, \bar{p}_{F}),
\end{align*}
hence $\bar{p}_F \in \BR^a_F(P_L)$.

Now, consider two distributions $p_{F} \in \BR^a_F(P_L)$ and $q_{F} \in \BR^a_F(P_L)$. Let $\bar{p}_{F} = \alpha p_{F} + (1 - \alpha) q_{F}$. Let $p^{*}_{L} = \argmin_{p_{L} \in P_{L}}U_{F}(p_{L},\bar{p}_{F})$. Then,
\begin{align*}
\min_{p_L \in P_L} U_F(p_L, \bar{p}_{F}) &= U_{F}(p^{*}_{L},\bar{p}_{F})\\
& = \alpha U_{F}(p^{*}_{L},p_{F}) + (1 - \alpha) U_{F}(p^{*}_{L},q_{F}) \\
& \geq \alpha \min_{p_L \in P_L} U_F(p_L, p_{F}) + (1 - \alpha) \min_{p_L \in P_L} U_F(p_L, q_{F}) \\
& = \min_{p_L \in P_L} U_F(p_L, p_{F}) \\
\end{align*}
Therefore, $\bar{p}_{F} \in \BR^a_F(P_L)$.
\end{proof}

\begin{definition}[Optimal Ambiguous Stackelberg Value]
    For a coupled Stackelberg game $\cG$ with ambiguous commitment and a tie-breaking rule $s$ for the followers, we define the \emph{optimal ambiguous Stackelberg value},
    \begin{align}
        \label{eq:ambiguous Stackelberg equilibrium}
        W^*(s) \coloneqq \sup_{P_L \in \cA(A_L)} W(s,  P_L).
    \end{align}
    We call $P_L^*$ with
    \begin{align*}
        W(s, P_L^*) \ge W^*(s) - \epsilon
    \end{align*}
    an \emph{$\epsilon$-approximately optimal, ambiguous Stackelberg strategy}.
\end{definition}
%The term ``StackEllsberg-equilibrium'' is a fusion of Heinrich Freiherr von Stackelberg, creator of sequential commitment games and Nazi-economist \citep{stolberg2010stackelberg}, and Daniel Ellsberg, who pupolarized the Ellsberg-paradox, a behavioral experiment in which ambiguity induces inconsistency with standard expected utility theory \citep{ellsberg1961risk}.
In contrast to our definition of optimal Stackelberg value, the optimal ambiguous Stackelberg value depends on the tie-breaking rule $s$. There is no such canonical choice as in the classical setting.

We have already hinted to the reader that closed, convex commitment sets are exhaustive. Hence, it is not necessary for the leader to consider the set of all subsets of $\Delta(A_L)$ as her action space. To formalize this statement, we introduce the notation $\co A$ for the closed, convex hull of a set $A$. Additionally, the extreme points of a convex set $A$ are $\ext A$.
\begin{lemma}
\label{lemma:closed, convex Commitment Sets are Exhaustive}
    Let $\cG$ be a coupled Stackelberg game.
    Let $P_L \subseteq \Delta(A_L)$ be arbitrary and define
    \begin{align*}
         \overline{\BR^a(P_L)} &\coloneqq \argmax_{ p_F \in \Delta(A_F)} \inf_{p_L \in P_L} U_F(p_L, p_F),\\
         \overline{W(s, P_L)} &\coloneqq \inf_{p_L \in P_L} \sum_{F \in \mathcal{F}}  U_{L_F}(p_L, s(\BR^a_F(P_L))).
    \end{align*}
    It holds
    \begin{align*}
        \overline{\BR^a_F(P_L)} = \BR^a(\co P_L) =  \BR^a(\ext \co P_L),
    \end{align*}
    and
    \begin{align*}
        \overline{W(s, P_L)} = W(s, \co P_L) = W(s, \ext \co P_L).
    \end{align*}
\end{lemma}
\begin{proof}
    We first argue that the following equation holds for arbitrary $P \subseteq \Delta(A_L)$
    and $c \in \reals^{|A_L|}$:
    \begin{align*}
         \inf_{p \in P} \langle p, c\rangle = \inf_{p \in \cl P} \langle p, c\rangle = \inf_{p \in \co P} \langle p, c\rangle = \min_{p \in \cl P} \langle p, c\rangle = \min_{p \in \co P} \langle p, c\rangle = \min_{p \in \ext \co P} \langle p, c\rangle.
    \end{align*}
    Note that the above equalities are equivalent to
    \begin{align*}
         \sup_{p \in P} \langle p, -c\rangle = \sup_{p \in \cl P} \langle p, -c\rangle = \sup_{p \in \co P} \langle p, -c\rangle = \max_{p \in \cl P} \langle p, -c\rangle = \max_{p \in \co P} \langle p, -c\rangle = \max_{p \in \ext \co P} \langle p, -c\rangle.
    \end{align*}
    The first three equalities follow from Proposition 2.2.1 in \citep{hiriart2004fundamentals}. The fourth and fifth equality hold because $\cl P$ (respectively $\co P$) is compact, since they are closed and subsets of a compact set $\Delta(A_L)$. Hence, the linear function from the compact set $\cl P$ (respectively $\co P$) into $\reals$ attains its maximum \citep[17.7.i]{schechter1996handbook}. The last equality holds following Proposition 2.4.6 in \citep{hiriart2004fundamentals}.

    The lemma follows immediately when observing that $U_F(p_L, p_F)$ and $\sum_{F \in \mathcal{F}}  U_{L_F}(p_L, s(\BR^a_F(P_L)))$ can be rewritten as $\langle p_L, c\rangle$ for some $c \in \reals^{|A_L|}$.
\end{proof}

Now that we have defined a model which equips the leader with ambiguity, it is natural to ask: do we recover the standard Stackelberg setting for an unambiguous commitment? We provide an affirmative answer to this question.
\begin{lemma}[Ambiguous Commitments Generalize Classical Commitments]
\label{lemma:reduction to Classical Stackelberg Games}
    Let $\cG$ be a coupled Stackelberg game. If the leader commits to a non-strictly ambiguous strategy, \ie, the leader commits to a singleton set $P_L = \{ q_L\} \subseteq \Delta(A_L)$, then
    \begin{align*}
        \BR^a_F(P_L) = \BR_F(q_L), \forall F \in \cF.
    \end{align*}
    For any tie-breaking rule $s$,
    \begin{align*}
        W(s, P_L) = V(s, q_L).
    \end{align*}
\end{lemma}
\begin{proof}
    The best response for follower $F \in \cF$ is given by
    \begin{align*}
        \BR^a_F(P_L) &\coloneqq \argmax_{p_F \in \Delta(A_F)} \min_{p_L \in \{ q_L\}} U_F(p_L,p_F)\\
        &= \argmax_{p_F \in \Delta(A_F)} U_F(q_L,p_F)\\
        &= \BR_F(q_L).
    \end{align*}
    The worst-case expected payoff for leader reduces to,
    \begin{align*}
        W(s, P_L) &= \min_{p_L \in P_L} \sum_{F \in \mathcal{F}}  U_{L_F}(p_L, s(\BR^a_F(P_L)))\\
        &= \sum_{F \in \mathcal{F}}  U_{L_F}(q_L, s(\BR^a_F(\{q_L\})))\\
        &= \sum_{F \in \mathcal{F}}  U_{L_F}(q_L, s(\BR(q_L))) = V_L(s, q_L).
    \end{align*}
    Note that since $\BR^a_F(P_L) = \BR_F(q_L)$, the tie-breaking rule selects the same element of the response set.
\end{proof}

\subsection{Implementing Ambiguity}
\label{sec:implementing ambiguity}
Just as how in classical game theory mixed strategies require randomization devices, ambiguous strategies require ``ambiguization'' devices. The exact nature of such devices is debatable. Scholars, such as \citep{beauchene2019ambiguous}, refer to Ellsberg's undetermined urn \citep{ellsberg1961risk}. In this case, the content of an urn is solely described by bounds on the number of colored balls in the urn. However, this leads to the question: who or what puts the balls into the urn? We interpret ``ambiguization'' as a service of a third party, which is opaque, \ie the party acts in an undetermined way, but is trusted, \ie the leader and followers believe that the third party will operate within its prescribed bounds (i.e. will only select a strategy in the ambiguous commitment set). This way we can justify the ambiguity-averse behavior of both the leader and followers. As we will show in this paper, ambiguity can be extremely valuable for a Stackelberg leader, motivating the idea that ``ambiguity-as-a-service," could exist.

The works of \citet{tillio2016design} and \citet{dütting2024ambiguouscontracts} solve the ambiguization problem in a slightly different way. They assume the leader themselves implements the ambiguization. But, the authors require that commitments are ``consistent'' which means that the expected leader payoffs for a fixed best maxmin response of the follower among all mixed strategies in the leader's ambiguous commitment are equal. In other words, the leader makes its ambiguous commitment credible by making itself indifferent to which mixed strategy from the commitment set is actually realized. In our coupled setting, determining as followers if a commitment is consistent in this way would involve complex reasoning and shared knowledge between multiple followers, and thus we find this restriction to be less meaningful.

Since we are considering potentially inconsistent ambiguous commitments and their worst-case expected payoff, we offer a slightly more general perspective. This way we can argue that our provided bounds to baselines (optimal classical Stackelberg value and the individualized Stackelberg value (ISV) (Definition~\ref{def:ISV})) hold for all attitudes of the leader against ambiguity. Furthermore, since the optimal value for a Stackelberg game with consistent ambiguous commitment is smaller than the optimal Stackelberg value for an arbitrary ambiguous commitment, Theorem~\ref{thm:single follower - no ambiguity needed} and Theorem~\ref{thm:Bounds to Ambiguous Stackelberg Value} still provide upper bounds on the former value. Even the poly-time algorithm for $2\times m$-games in Section~\ref{sec:poly-time} can be adapted to provide optimal consistent ambiguous commitments (Remark~\ref{rem:consistent}). For a broader discussion of the nature and origin of ambiguity in games we refer the reader to, \eg \citep{binmore2007rational, sass2013ellsberg}.

\section{Ambiguity Advantage} \label{sec:ambiguity}
Coupled Stackelberg games with ambiguity generalize classical, coupled Stackelberg games. But is the power of ambiguity at all helpful? Are there settings where a leader will choose to commit to an ambiguous set rather than a precise distribution? Towards our positive answer for this question, we introduce the notion of an ambiguity gap.
\begin{definition}[Ambiguity Gap]
\label{def:ambiguity gap}
    Let $\cG$ be a coupled Stackelberg game. Let $\Bar{s}$ be the strong tie-breaking rule. If $V^*(\Bar{s}) \neq 0$ and $W^*(s) \neq 0$ for a tie-breaking rule $s$, then the \emph{ambiguity gap} is defined by
    \begin{align*}
        G(s) \coloneqq \frac{|W^*(s)|^{\sgn(W^*(s))}}{|V^*(\Bar{s})|^{\sgn(V^*(\Bar{s}))}}.
    \end{align*}
    If $G > 1$, we say that the leader has an \emph{ambiguity-advantage} in the Stackelberg game.
\end{definition}
Obviously, $G \ge 1$ (Lemma~\ref{lemma:reduction to Classical Stackelberg Games}). Having a multiplicative definition of ambiguity gap makes the quantity invariant to rescaling of utilities in a coupled Stackelberg game $\cG$. This definition might look overly complicated---note that, if both $W^{*}(s)$ and $V^{*}(s)$ are positive, we recover the definition of an ``ambiguity gap'' as given in \citep{dütting2024ambiguouscontracts}. However, we require the sign handling to work with potentially negative values, as restricting to positive leader utilities is not without loss of generality. We show in Lemma~\ref{lem:positive_ub} that, in zero-sum games where the leader has only positive utilities, the ambiguity gap is bounded above by $k$, the number of followers. But there exist two follower zero-sum games with non-positive utilities for the leader in which the ambiguity gap is unbounded (Theorem~\ref{thm:amb_gap_zs}). Generally, note that $W^*(s) > V^*(\Bar{s})$ implies $G(s) > 1$.

\subsection{Single-follower Games}
It is not immediately clear what role ambiguity plays; adding more actions to the ambiguous set may cause the follower(s) to change their behavior in favorable ways, but it may also decreases the worst case utility of the leader over this set. 

Indeed, in the case of a single follower, we show that there is no reason to use ambiguity. Intuitively, any behavior that that the leader can incentivize for this follower with an ambiguous set, she can also incentivize with a single distribution, while attaining at least as good worst-case payoff for herself. However, in the case of multiple followers, we show that the ambiguity gap can be unbounded.

%\begin{itemize}

  %  \item A similar statement is true for so-called Ellsberg-equilibria, which are extended Nash-equilibria for imprecise mixed strategies of agents. For two-player games Ellsberg and Nash-equilibria are the same [The Strategic Use of Ambiguity]. What the authors then show is that in three player games there are more Ellsberg-equilibria which are not in the support of the Nash-equilibria.
%\end{itemize}

\begin{theorem}
\label{thm:single follower - no ambiguity needed}
    Let $\cG = (A_L, A_F, u_F, u_L)$ be an ambiguous Stackelberg game with a single follower. For all tie-breaking rules $s$ it holds $G(s) = 1$.
\end{theorem}
\begin{proof}
    To show $G(s) \le 1$, we prove the following statement: for all tie-breaking rules $s$ and for all closed, convex $P_L \subseteq \Delta(A_L)$,
    \begin{align*}
        W(s, P_L)\coloneqq \min_{p_L \in P_L}U_L(p_L, s(\BR^a(P_L))) \le \max_{p_L \in \Delta(A_L)}U_L(p_L, \Bar{s}(\BR(p_L))] \eqqcolon V^*(\Bar{s}).
    \end{align*}
    
    Clearly, $\emptyset \neq \BR^a(P_L) \subseteq \Delta(A_F)$. We now show that there exists $d \in P_L$ such that $\BR^a(P_L) \subseteq \BR(d)$. To see this, let us focus on the following zero-sum game. Let the follower be the maximizer with action space $\Delta(A_F)$. Let the leader be the minimizer with action space $P_L$. The payoff is defined through $U_F$, concretely $u\colon \Delta(A_F) \times P_L \rightarrow \reals$ with $u(p_F, p_L) \coloneqq U_F(p_L, p_F)$. Hence, $\BR^a(P_L)$ is the set of the follower's maxmin-strategies in this zero-sum game. In particular, there exists a Nash equilibrium $p^*_F \in \BR^a(P_L), p^*_L \in P_L$ such that $p^*_F$ is a maxmin response to $P_L$ and $p^*_L$ is a minmax response to $\Delta(A_F)$. We choose $d = p^*_L$ which gives $\BR^a(P_L) \subseteq \BR(d)$. Finally,
    \begin{align*}
        \min_{p_L \in P_L}U_L(p_L, s(\BR^a(P_L))) &\le U_L(d, s(\BR^a(P_L)))\\
        &\le U_L(d, \Bar{s}(\BR^a(P_L)))\\
        &\le U_L(d, \Bar{s}(\BR(d)))\\
        &\le V^*(\Bar{s}).
    \end{align*}
\end{proof}

\subsection{Coupled Games}

In contrast to single-follower games, we show that in coupled Stackelberg games, committing to an ambiguous set can be strictly better than committing to a single distribution. Below, we prove the existence of an ambiguity gap even in the simple setting of one leader and two followers with $n=m=2$. We give an example where the gap is unbounded even in a zero-sum game.
%Our initial benchmark begin by considering settings in which the leader must play the same action against all other agents, and may choose to play precisely or ambiguously. 

\begin{theorem} 
For any $C \in \reals$, there exists a coupled, zero-sum Stackelberg game with 2 followers in which the ambiguity gap $G(s) > C$ for all tie-breaking rules $s$. \label{thm:amb_gap_zs}
\end{theorem}
\begin{proof}
We will give a proof by example. The game we define here consists of two simultaneously played zero-sum games between a leader and two followers. We show that in this game there is an imprecise mixed strategy which dominates all precise mixed strategies.

Let $D > 2C$. Let the outcome matrix for follower $F_1$ and $F_2$ be given as in the following Tables~\ref{tab:parasol-umbrella-F1} and \ref{tab:parasol-umbrella-F2}.
\begin{table}[h]
    \centering
    \begin{minipage}{0.45\textwidth}
        \centering
        \begin{tabular}{c|c c}
             $F_1$ & $b_1$ & $b_2$\\ \hline
            $a_L = 0$ & D &  1\\
            $a_L = 1$ & 0 & 1
        \end{tabular}
        \caption{Follower $F_1$'s payoffs.}
        \label{tab:parasol-umbrella-F1}
    \end{minipage}%
    \hfill
    \begin{minipage}{0.45\textwidth}
        \centering
        \begin{tabular}{c|c c}
             $F_2$ & $c_1$ & $c_2$ \\ \hline
            $a_L = 0$ & 1 & 0\\
            $a_L = 1$ & 1 & D
        \end{tabular}
        \caption{Follower $F_2$'s payoffs.}
        \label{tab:parasol-umbrella-F2}
    \end{minipage}
\end{table}

The leader incurs the sum of the negative utilities of the followers.

First, we consider the cases, where the leader gives precise mixed strategies $p \in [0,1]$. Consider the outcomes for follower $F_1$:
\begin{enumerate}
    \item For $p < 1-\frac{1}{D}$, $F_1$ plays $b_1$, hence the payoff for leader will be $-(1-p)D$.
    \item For $p = 1-\frac{1}{D}$, $F_1$ plays $b_1$ or $b_2$, hence the payoff for leader will be $= -1$.
    \item For $p > 1-\frac{1}{D}$, $F_1$ will play $b_2$, hence the payoff for leader will be $= -1$.
\end{enumerate}
And consider the outcomes for follower $F_2$:
\begin{enumerate}
    \item For $p < \frac{1}{D}$, $F_2$ will play $c_1$, hence the payoff for leader will be $= -1$.
    \item For $p = \frac{1}{D}$, $F_2$ will play $c_1$ or $c_2$, hence the payoff for leader will be $= -1$.
    \item For $p > \frac{1}{D}$, $F_2$ will play $c_2$, hence the payoff for leader will be $-pD $.
\end{enumerate}
It is easy to see that $\frac{1}{D} < 1 - \frac{1}{D}$ for $D > 2$. We summarize the outcomes for both followers simultaneously:
\begin{enumerate}
    \item For $p \le \frac{1}{D}$, leader's total payoff is $-1 - (1-p) D \le - D$.
    \item For $\frac{1}{D} < p \le 1-\frac{1}{D}$, leader's total payoff is $- p D - (1-p) D = - D$.
    \item For $1-\frac{1}{D} < p$, leader's total payoff is $- p D - 1 \le - D$.
\end{enumerate}

In comparison, a fully imprecise mixed strategy, \ie vacuous set of all probability distributions $P_L = \Delta(A_L)$, yields a payoff $= -2$. The reason for this is that the optimal maxmin strategy against ambiguous commitment $P_L$ is $b_2$ (respectively $c_1$). The intuition is that the followers play the safe option. In each sub-game the worst-case payoff is therefore $-1$, which sum up to $-2$. This gives the ambiguity gap
\begin{align*}
    G(s) = \frac{|W^*(s)|^{\sgn(W^*(s))}}{|V^*(\Bar{s})|^{\sgn(W^*(\Bar{s}))}} = \frac{|V^*(\bar{s})|}{|W^*({s})|} \ge \frac{D}{2} > C.
\end{align*}
\end{proof}

\section{Ambiguous Coupling Advantage}
\label{sec:coupling}
Now that we have established how large of an advantage ambiguous commitment   can be, we can ask for an even stronger benchmark: what if the leader has the choice between committing to the same ambiguous set for all followers, or committing to a different distribution against each follower individually? Note that the latter case is, by Theorem~\ref{thm:single follower - no ambiguity needed}, exactly the same as allowing them to commit to ambiguous sets against each follower individually. Might she commit to a single ambiguous set for all followers, \emph{even if} she could instead tailor her actions to each follower in every individual game? We define a new benchmark called the individualized Stackelberg value to compete with.
\begin{definition}[Individualized Stackelberg Value (ISV)]
\label{def:ISV}
Let $(\cG_F)_{F \in \cF}$ be a family of decoupled, classical Stackelberg games corresponding to a coupled Stackelberg game $\cG$. Let $\Bar{s}$ be the strong tie-breaking rule. The \emph{individualized Stackelberg value (ISV)} of $(\cG_F)_{F \in \cF}$ is the sum of optimal Stackelberg values $V^*_F(\Bar{s})$ for each game:
\begin{align*}
  ISV \coloneqq \sum_{F \in \mathcal{F}}V^*_F(\Bar{s})  
\end{align*}
\end{definition}

If the leader is restricted to unambiguous commitments, there is no advantage in coupling multiple followers. This is because any mixed strategy which is broadcast to coupled followers could have been sent to each individual follower in a decoupled setting and lead to the same response outcome. This intuition is formalized in Lemma~\ref{lem:ISV_lb}. However, we will show that, with ambiguity, the story is different.

\begin{definition}[Ambiguous Coupling Gap]
\label{def:ambiguous coupling gap}
    Let $\cG$ be a coupled Stackelberg game. If $ISV > 0$ and $W^*(s) > 0$\footnote{For the sake of simplicity we kept this definition only for positive values. In contrast to the discussion around Definition~\ref{def:ambiguity gap}, we have not found any curious abnormalities for non-positive utility games.} for a tie-breaking rule $s$, then the \emph{Ambiguous Coupling Gap} is defined by
    \begin{align*}
        C(s) \coloneqq \frac{W^*(s)}{ISV}.
    \end{align*}
    If $C(s) > 1$, we say that the leader has a \emph{coupling advantage} in the Stackelberg game.
\end{definition}
In this section we show that it is impossible to beat the ISV benchmark when all follower games are zero-sum, but also that, perhaps surprisingly, it \emph{is} possible to beat this benchmark in general. Thus, even when a leader is engaging in \emph{separate and possibly unrelated} strategic interactions, it may make sense for her to correlate her actions between these games and commit to a single ambiguous set. Essentially, it is sometimes worth giving up individualization in order to use ambiguity.

\subsection{Zero-Sum Games}   

First, we show that the benefit of coupling and ambiguity in tandem is not present in zero-sum games. By Theorem~\ref{thm:amb_gap_zs}, ambiguity can be a beneficial tool for a leader in an environment where she must naturally commit to the same strategy in multiple zero-sum games. However, this benefit does not extend to the decoupled setting--if the leader is given the opportunity to commit separately in each of these zero-sum games, she always should. And by Theorem~\ref{thm:single follower - no ambiguity needed}, these decoupled commitments need not be ambiguous.  

     \begin{theorem}[No Coupling Gap in Coupled, Zero-Sum Stackelberg Games]
    \label{thm:Bounds to Ambiguous Stackelberg Value}
        Let $\cG$ be a coupled, zero-sum Stackelberg game with $u_F = -u_{L_F}$ for all $F \in \cF$. For all tie-breaking rules $s$,
        \begin{align*}
            C(s) \le 1.
        \end{align*}
    \end{theorem}
    \begin{proof}
        We show that $W^*(s) \le ISV$ for all tie-breaking rules $s$.
        Let $P^* \subseteq \Delta(A_L)$ be the optimal ambiguous commitment in the coupled, zero-sum Stackelberg game $\cG$. For all $q_L \in P^*$ and all tie-breaking rules $s$,
        \begin{align*}
            U_F(q_L, s(\BR_F^a(\mathcal{P}^*))) &\ge \min_{p_L \in P^*} U_F(p_L, s(\BR_F^a(\mathcal{P}^*)))\\
            &= \max_{p_F \in \Delta(A_F)} \min_{p_L \in P^*} U_F(p_L, p_F)\\
             &=\min_{p_L \in \cP^*} \max_{p_F \in \Delta(A_F)}  U_F(p_L, p_F)\\
             &\ge\min_{p_L \in \Delta(A_L)} \max_{p_F \in \Delta(A_F)}  U_F(p_L, p_F)\\
        \end{align*}
        by Von Neumann's minmax theorem. Hence, for any tie-breaking rule $s$,
        \begin{align*}
            W^*(s) &\coloneqq \min_{p_L \in \mathcal{P}^*} \sum_{F \in \mathcal{F}} U_{L_F}(p_L, s(\BR^a_F(\mathcal{P}^*)))\\
            &= - \max_{p_L \in \mathcal{P}^*} \sum_{F \in \mathcal{F}} U_{F}(p_L, s(\BR^a_F(\mathcal{P}^*)))\\
            &\le - \sum_{F \in \mathcal{F}} \min_{p_L \in \Delta(A_L)} \max_{p_F \in \Delta(A_F)}  U_F(p_L, p_F)\\
            &= \sum_{F \in \mathcal{F}} - \min_{p_L \in \Delta(A_L)} u_{F}(p_L, \Bar{s}(\BR_F(p_L)))\\
            &= \sum_{F \in \mathcal{F}} \max_{p_L \in \Delta(A_L)} u_{L_F}(p_L, \Bar{s}(\BR_F(p_L)))\\
            &= \sum_{F \in \mathcal{F}} V^*_F(\Bar{s}).
        \end{align*}
    \end{proof}
    We can show that the bound is tight, in the sense that there exists a coupled zero-sum game for which $C(s) = 1$ (see Table~\ref{tab:parasol-umbrella-F1} and Table~\ref{tab:parasol-umbrella-F2}).

\begin{remark}
    \label{remark:Necessary Condition for Ambiguity Gap}
    Note that Theorem~\ref{thm:Bounds to Ambiguous Stackelberg Value} provides a necessary condition for an ambiguity gap. Let $\cG$ be a coupled, zero-sum Stackelberg game. We denote the set of optimal, classical Stackelberg strategies in the corresponding decoupled, zero-sum game as $p^*_L(F) \coloneqq \{ p \in \Delta(A_L)\colon U_F(p, s(BR(p))) = V^*\}$. If $\bigcap_{F \in \mathcal{F}} p^*_L(F) \neq \emptyset$ then there exists $p^* \in \bigcap_{F \in \mathcal{F}} p^*_L(F)$ such that $p^*$ is an optimal, classical Stackelberg strategy in the coupled, zero-sum Stackelberg game. Hence, $ISV = V^*(\Bar{s})$ which implies $G(s) = C(s) \le 0$.
\end{remark}

%\begin{observation} For all game sets $\mathcal{G}$, $ISV(\mathcal{G}) \leq GSV(\mathcal{G})$
%\end{observation}

\subsection{General-Sum Games}
One might expect that, when playing $k$ entirely unrelated Stackelberg games against $k$ different followers, the best thing for the leader to do is to optimize her utility in each game separately. But it turns out that, when the followers are ambiguity-averse, there is sometimes a better option: instead of committing separately in each game, artificially \emph{couple} the games by committing to the same ambiguous commitment set in each game.\footnote{One can define a new action index for the leader and a mapping associating this index with its respective action in each game} There are decoupled Stackelberg game settings where a leader can do better with this strategy--even unboundedly better. 
%\rab{why does it work. Because using ambiguous commitments the leader can provoke reactions and simultaneously hedge against the worst-case scenarios. equipped with only mixed commitments, the leader can provoke the same reactions but cant necessarily hedge against the worst-case outcomes.}

\begin{theorem}
For any $C \in \reals$, there exists a coupled Stackelberg game with 2 followers in which the ambiguous coupling gap $C(s) > C$ for all tie-breaking rules $s$. \label{thm:coupling_gap}
\end{theorem}
\begin{proof}
We provide a proof by example.
Let $D > 2C$. Consider the coupled Stackelberg game with two followers as summarized in Tables~\ref{tab:F1}, \ref{tab:L1}, \ref{tab:F2} and \ref{tab:L2}. Let us call the game against the first follower $G_1$ and the game against the second follower $G_2$.

First, we calculate the leader's payoff for a fully ambiguous commitment. Then we compare this payoff with an upper bound on the ISV. It turns out that we can arbitrarily scale the ISV-gap via the parameter $D$.

Let the leader commit to full ambiguity. The maxmin response of the follower in $G_1$ is $a_{2}$, and of the follower in $G_2$ is $b_{1}$. Thus against full ambiguity, the leader will get payoff $1$. %Note that the payoffs for leader in $G_1$ and $G_2$ hedge.

By contrast, let us consider the classical Stackelberg value of each game separately. In $G_1$, to incentivize $a_{2}$, the leader must play her second action at least with $\frac{D-1}{D}$ probability. This means that the leader's payoff is upper bounded at $\frac{1}{D}$. The same is true of $G_2$ by symmetry. Thus, $ISV \le \frac{2}{D}$, hence,
\begin{align*}
    C(s) = \frac{W^*(s)}{ISV} \ge \frac{1}{\frac{2}{D}} > C.
\end{align*}

\begin{table}[h]
    \centering
    \begin{minipage}{0.45\textwidth}
        \centering
        \begin{tabular}{c|c c}
             $F_1$ & $a_1$ & $a_2$\\ \hline
            $a_L = 0$ & $D$ &  1\\
            $a_L = 1$ & 0 & 1
        \end{tabular}
        \caption{Follower $F_1$'s payoffs.}
        \label{tab:F1}
    \end{minipage}%
    \hfill
    \begin{minipage}{0.45\textwidth}
        \centering
        \begin{tabular}{c|c c}
             $L_1$ & $a_1$ & $a_2$ \\ \hline
            $a_L = 0$ & 0 & 1\\
            $a_L = 1$ & 0 & 0
        \end{tabular}
        \caption{Leader's payoff against $F_1$.}
        \label{tab:L1}
    \end{minipage}
    \vspace{0.5cm} % Add some space between the rows of tables
    \begin{minipage}{0.45\textwidth}
        \centering
        \begin{tabular}{c|c c}
             $F_2$ & $b_1$ & $b_2$ \\ \hline
            $a_L = 0$ & 1 & 0\\
            $a_L = 1$ & 1 & $D$
        \end{tabular}
        \caption{Follower $F_2$'s payoffs.}
        \label{tab:F2}
    \end{minipage}%
    \hfill
    \begin{minipage}{0.45\textwidth}
        \centering
        \begin{tabular}{c|c c}
             $L_1$ & $b_1$ & $b_2$ \\ \hline
            $a_L = 0$ & 0 & 0\\
            $a_L = 1$ & 1 & 0
        \end{tabular}
        \caption{Leader's payoff against $F_2$.}
        \label{tab:L2}
    \end{minipage}
\end{table}

\end{proof}

\section{A Poly-time Algorithm for $2 \times m$-Stackelberg Games} \label{sec:poly-time}
% The computation of optimal ambiguous commitments requires a demanding optimization over a partially continuous search space, which is the set of all closed, convex sets in $\Delta(A_L)$. In contrast to the hardness of a general sum $k \times m$-game, it turns out that the complexity collapses in the case of $2 \times m$-games. 
In Sections~\ref{sec:ambiguity} and \ref{sec:coupling}, we show the power of ambiguous commitment. In Appendix~\ref{app:structure}, we show that the optimal ambiguous commitment is not always of a simple form, and can have extreme points which are not pure strategies. Because of this, it is not a-priori clear whether there is even a finite-time algorithm for computing the optimal ambiguous commitment.\footnote{If the optimal commitment strategy always corresponded to the convex hull of a subset of pure actions, there would be an exponential but finite time enumeration algorithm. But there are uncountably many mixed strategies, so there is no clear brute force algorithm based on enumeration.} In this section we provide a finite, polynomial time algorithm (polynomial in $k$ and $m$) in the special case in which the leader has $n=2$ actions. In the next section, we show that it is unlikely that this algorithm can be extended substantially, as the general case (for general $m$) is NP-hard. 

\begin{definition}[Coupled $2 \times m$-Stackelberg Game]
    \label{def: coupled 2x m Stackelberg Game}
    Let $\cG$ be a coupled Stackelberg game. If $|A_L| = 2$ we call $\cG$ a \emph{coupled $2 \times m$-Stackelberg game}.
\end{definition}

A key simplification in coupled $2 \times m$-Stackelberg games with ambiguous commitments is that a mixed strategy of the leader can be described by a single number (the probability that the leader places on action $1$), and the convex hull of a set of mixed strategies is thus just an interval, which can be defined by its two endpoints.\footnote{This is a consequence of the closed, compact, convex nature of ambiguous commitments (Lemma~\ref{lemma:closed, convex Commitment Sets are Exhaustive}). The extreme points of the set exist and define the set (\citep[Proposition 2.4.6]{hiriart2004fundamentals}).} Hence, the main idea for the computation is straightforward: we restrict the search space of potential ambiguous commitments. First, we provide a characterization of the best response landscape for a follower given an unambiguous commitment by the leader. We then provide an analogous characterization of maxmin responses to ambiguous commitments made by the leader. Finally, we exploit the structure by providing a polynomially small set $N$ of potential endpoints for ambiguous commitments which exhaustively induce all potential maxmin responses (Lemma~\ref{lemma:N2 is exhaustive}). The ($\epsilon$-approximate) optimal commitment is provably one of those commitments with endpoints in $N$. Hence, the algorithm brute-forces over the simplified search space for the best worst-case expected payoff.

Let $\cG$ be a coupled $2 \times m$-Stackelberg game with $u_{L_F}, u_F \in \cM( 2 \times m)$ for all $F \in \cF$. \textbf{Throughout the entire section we assume that each follower has no weakly dominated strategy in his payoff matrix $u_F$.} This is a mild assumption. In cases where the follower has to tie-break, weakly dominated strategies could change their behavior. In cases where the response of follower involves no tie-breaking the assumption is without loss of generality. We make it to simplify our treatment of tie breaking. Let us first focus on a single follower $F \in \cF$. We explicitly write down the payoff matrix for this follower with action space $A_F = \{ a_1, a_2, \ldots , a_n\}$ and the leader's action space $A_L = \{ i, ii\}$,
\begin{align*}
    u_F = \begin{pmatrix}
        & a_1 & a_2 & \ldots & a_n \\
       i & w_1 & w_{2} & \ldots & w_{n} \\
       ii & v_1 & v_{2} & \ldots & v_{n}
    \end{pmatrix}
\end{align*}
We use, with slight abuse of notation, the shorthand $p \in [0,1]$ if the leader plays the mixed strategy $(1-p) i + p \ ii$. We define the linear utility function in the leader's strategy $p$ for every action $a \in A_F$,
\begin{align*}
    U_j(p)\coloneqq (1-p) w_j + p v_j= u_j + p (v_j - w_j).
\end{align*}

\begin{figure}
    \centering
    \def\svgwidth{0.45\columnwidth}
    {\footnotesize
    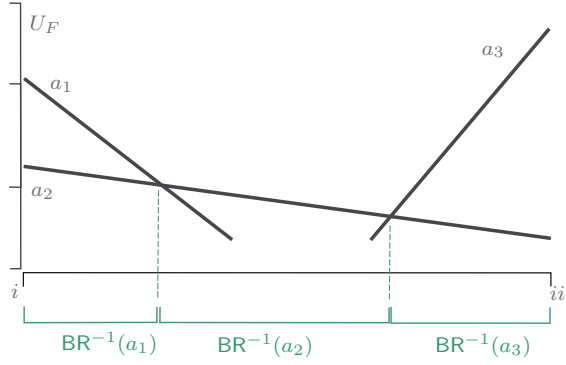
    }
    \caption{Best Response Landscape for an example $2 \times 3$-Game.}
    \label{fig:best response landscape}
\end{figure}

Now, let us define
\begin{align*}
    \BR^{-1}\colon A_F \rightarrow 2^{\Delta(A_L)}, \quad \BR^{-1}(a) \coloneqq \{p \in \Delta(A_L) \colon a \in \BR(p) \}.
\end{align*}
There exists an order on slopes, and this order determines the structure of the best response landscape. For an illustration of the best response landscape, see Figure~\ref{fig:best response landscape}.

\begin{lemma}[Best Response Landscape]
\label{lemma:best response landscape}
    Let $\cG$ be a coupled $2 \times m$-Stackelberg game with a single follower. Let the follower have no weakly dominated strategy in $u_F$. There exists an order structure on the actions $A_F$ such that,
    \begin{align*}
        \BR^{-1}(a_1) = [0, \mu_1], \BR^{-1}(a_2) = [\mu_1, \mu_2], \ldots, \BR^{-1}(a_{n}) = [\mu_{n-1}, 1],
    \end{align*}
    where $\mu_j$ such that $U_{j}(\mu_j) = U_{j+1}(\mu_{j})$, and
    \begin{align*}
        v_1 - w_1 < v_2 - w_2 < \ldots v_n - w_n.
    \end{align*}
\end{lemma}
\begin{proof}
    Lemma~\ref{lemma: Best response sets are closed and convex} and the assumption that there are no weakly dominated strategies for the follower show
    that for each action $a \in A_F$ a best response interval $\BR^{-1}(a) \subseteq [0,1]$ exists.
    
    These intervals only overlap on singletons: note that $q \in \BR^{-1}(a_j)$ if and only if $U_j(q) \ge U_i(q)$ for all $i \in [n]$. Furthermore, if $q \in \BR^{-1}(a_j) \cap \BR^{-1}(a_{i})$ then $U_j(q) = U_i(q)$. There does not exist $q, q' \in [0,1], q \neq q'$ such that there exist $i, j \in [n], i\neq j$ with $U_i(q) = U_j(q)$ and $U_i(q') = U_j(q')$, because $U_i$ and $U_j$ are linear functions, and if they coincide in two points then $U_j = U_i$, hence $i = j$.

    Furthermore, for every $q \in [0,1]$ a best response exists. Hence, without loss of generality these intervals can be ordered and fill the entire action space $[0,1]$.  In particular, $U_j(q) = U_{j+1}(q)$ for $q = \mu_j$.

    It remains to show that the order of the best response intervals is equivalent to the order of slopes
    \begin{align*}
        v_1 - w_1 < v_2 - w_2 < \ldots v_n - w_n.
    \end{align*}
    The argument is given by contradiction. Assume there exists an $j \in [n-1]$ such that $v_j - w_j \ge v_{j+1} - w_{j+1}$. By definition $U_j(q) \ge U_{j+1}(q)$ for all $q \in [\mu_{j-1}, \mu_j]$ (where $\mu_0 = 0$). In particular, $U_j(\mu_j) = U_{j+1}(\mu_j)$. But then if $v_j - w_j \ge v_{j+1} - w_{j+1}$, for $p \in [\mu_j, \mu_{j+1}]$ (where $\mu_n = 1$),
    \begin{align*}
        U_j(p)&\coloneqq u_j + p (v_j - w_j)\\
        &= u_j + \mu_j (v_j - w_j) + (p - \mu_j) (v_j - w_j)\\
        &= U_j(\mu_j) + (p - \mu_j) (v_j - w_j)\\
        &= U_{j+1}(\mu_j) + (p - \mu_j) (v_j - w_j)\\
        &\ge U_{j+1}(\mu_j) + (p - \mu_j) v_{j+1} - w_{j+1} = U_{j+1}(p),
    \end{align*}
    since $p - \mu_j \ge 0$, which is in contradiction to the definition of the best response set $\BR^{-1}(a_{j+1}) = [\mu_j, \mu_{j+1}]$.
\end{proof}
We define the action sets $A^- \coloneqq \{a_j \in A_F\colon w_j > v_j \}$, $A^+ \coloneqq \{a_j \in A_F\colon w_j < v_j \}$ (and $A^= \coloneqq \{a_j \in A_F\colon w_j = v_j \}$). The signs represent the slope of $U_j$. The sets collect the follower's actions which are less likely to be chosen if the leader increases the weight on its second action $ii$ ($A^-$), the ones which are more likely to be chosen if the leader increases the weight on its second action $ii$ ($A^+$) and all constant utility actions $A^=$.

\begin{figure}
\captionsetup[subfigure]{labelformat=empty}
\begin{subfigure}[h]{0.48\linewidth}
    \centering
    \def\svgwidth{0.99\columnwidth}
    {\footnotesize
    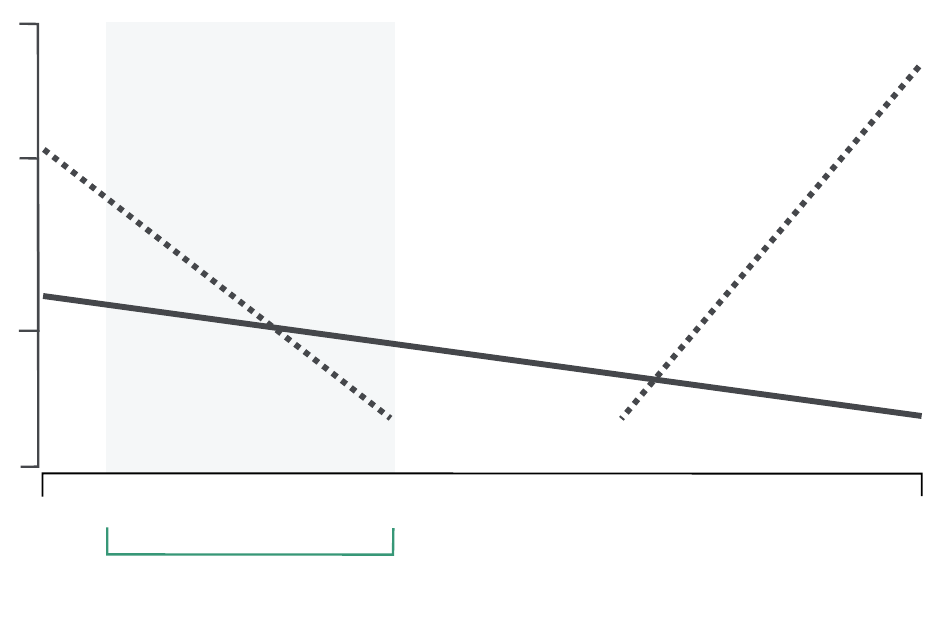
    }
    \caption{Maxmin-best response for an ambiguous commitment left of the point of slope sign change $\mu^{\pm}$ for an example $2 \times 3$-game (Case (a) in Lemma~\ref{lemma:maxmin best responses -- empty equal action}).}
    \label{fig:left of slope sign change}
\end{subfigure}
\hfill
\begin{subfigure}[h]{0.48\linewidth}
    \centering
    \def\svgwidth{0.99\columnwidth}
    {\footnotesize
    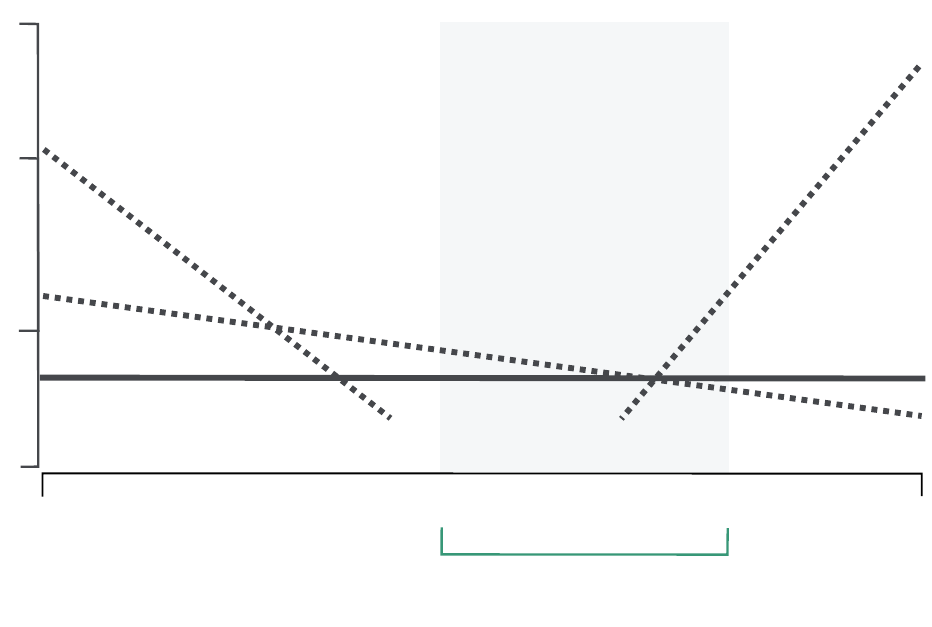
    }
    \caption{Maxmin-best response for an ambiguous commitment including the point of slope sign change $\mu^{\pm}$ for an example $2 \times 3$-game (Case (e) in Lemma~\ref{lemma:maxmin best responses -- empty equal action}).}
    \label{fig:over slope sign change}
\end{subfigure}
\caption{Maxmin-best responses for ambiguous commitment on examplary $2\times 3$-game. In this example, $a_2 = a_x$ and $a_3 = a_y$. The intuition for the maxmin responses is relatively straight forward. The lower $\underline{p}$ and upper endpoint $\overline{p}$ of the ambiguous commitment define a region (grey) of potential expected payoff for follower. Then, the follower chooses the action or the convex combination of actions (solid black line) which maximizes the worst case payoff (green mark) within this region. Note that in this plot the expected payoff of a convex combination of two actions (\eg $t_\nu$) shows up as a line crossing the intersection of the two actions and staying bounded between the two actions.}
\label{fig:maxmin best responses -- empty equal action}
\end{figure}

As we are dealing with ambiguous commitments by the leader, we reiterate the analogous characterization for maxmin responses of the followers. We distinguish between two main cases: (a) $A^= = \emptyset$ and (b) $A^= \neq \emptyset$. For now, we assume that $A^= =\emptyset$. Hence, there exist an action pair $a_x \in A^-$ and $a_y \in A^+$ which touch each other in $\mu^{\pm}$, \ie, $U_x(\mu^{\pm}) = U_y(\mu^{\pm})$. Figure~\ref{fig:maxmin best responses -- empty equal action} depicts the setting and provides some intuition. In particular, it illustrates the cases (a) and (e) of the following lemma.

\begin{figure}
    \centering
    \def\svgwidth{0.45\columnwidth}
    {\footnotesize
    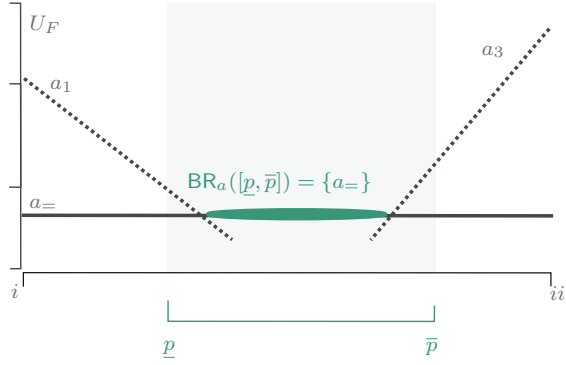
    }
    \caption{Maxmin-best response for an ambiguous commitment including the action of slope sign change $a_=$ for an example $2 \times 3$-game (Case (c) in Lemma~\ref{lemma:maxmin best responses -- non-empty equal action}). As in Figure~\ref{fig:maxmin best responses -- empty equal action} follower's maxmin response (solid line) maximizes the worst case expected payoff (green mark) within the region defined through the endpoints of the ambiguous interval (grey).}
    \label{fig:over no sign change}
\end{figure}

\begin{lemma}
\label{lemma:maxmin best responses -- empty equal action}
    Let $\cG$ be a coupled $2 \times m$-Stackelberg game with a single follower. Let the follower have no weakly dominated strategy in $u_F$. Assume that $A^= =\emptyset$. Define the point of slope sign change as $\mu^\pm$. The action left of the slope sign change is $a_x$, the action right of it is $a_y$.
    Let
    \begin{align*}
        \nu\coloneqq \frac{1}{\frac{v_y - w_y}{w_x - v_x} + 1}.
    \end{align*}
    We define the action $t_p \coloneqq (1-p)\cdot a_x + p \cdot a_y$
    \begin{enumerate}[(a)]
        \item If $ \underline{p} \le \overline{p} < \mu^\pm$, then $\BR^a(\upop) = \BR(\overline{p})$.
        \item If $\underline{p} < \overline{p} = \mu^\pm$, then $\BR^a(\upop) = \Delta(\{a_x, t_\nu\}) = \{t_p \colon p \in [0, \nu] \}$.
        \item If $\underline{p} = \overline{p} = \mu^\pm$, then $\BR^a(\upop) = \Delta(\{a_x,a_y\}) = \BR(\mu^\pm)= \{t_p \colon p \in [0, 1] \}$.
        \item If $ \mu^\pm = \underline{p} \le \overline{p}$, then $\BR^a(\upop) = \Delta(\{t_\nu, a_y\}) = \{t_p \colon p \in [\nu, 1] \}$.
        \item If $\underline{p} <  \mu^\pm <  \overline{p}$, then $\BR^a(\upop) = \{ t_\nu\}$.
        \item If $\mu^\pm < \underline{p} \le \overline{p} $, then $\BR^a(\upop) = \BR(\underline{p})$.
    \end{enumerate}
\end{lemma}
The proof of this lemma is tedious but not very insightful---the reder can find it in the Appendix, see Proof~\ref{proof:maxmin best responses -- empty equal action}. In the next lemma, we assume that there exists a constant payoff action for follower, called $a_=$. Note that $U_{a_=}$ has slope $0$. The case (a) (respectively (b)) resembles the case (a) (respectively (f)) of the previous lemma and has the same intuition. Case (c) is new and is a consequence of $A^= \neq \emptyset$. Figure~\ref{fig:over no sign change} provides an illustrative explanation for this case.

\begin{lemma}
\label{lemma:maxmin best responses -- non-empty equal action}
    Let $\cG$ be a coupled $2 \times m$-Stackelberg game with a single follower. Let the follower have no weakly dominated strategy in $u_F$. We assume $\cA^= = \{ a_=\}$ (without loss of generality it is a singleton). We define $\mu^{-=}$ such that $U_{x}(\mu^{-=}) = U_{a_=}(\mu^{-=})$ and $\mu^{=+}$ such that $U_{y}(\mu^{=+}) = U_{a_=}(\mu^{=+})$. The action left of the slope sign change is $a_x$, the action right of it is $a_y$.
    \begin{enumerate}[(a)]
        \item If $ \underline{p} \le \overline{p} < \mu_{=+}$, then $\BR^a(\upop) = \BR(\overline{p})$.
        \item If $\mu_{-=} < \underline{p} \le \overline{p} $, then $\BR^a(\upop) = \BR(\underline{p})$.
        \item If $\underline{p} \le  \mu_{-=} < \mu_{=+} \le \overline{p}$, then $\BR^a(\upop) = \{a_=\}$.
    \end{enumerate}
\end{lemma}
Due to its technical, but not insightful structure, the proof is in the appendix Proof~\ref{proof:maxmin best responses -- non-empty equal action}. With the two preceding lemmas we can describe all evocable maxmin responses by an ambiguous commitment for a single follower. In fact, the statements are sufficient to describe the maxmin responses even for multiple followers.

\begin{figure}
    \centering
    \def\svgwidth{0.55\columnwidth}
    {\footnotesize
    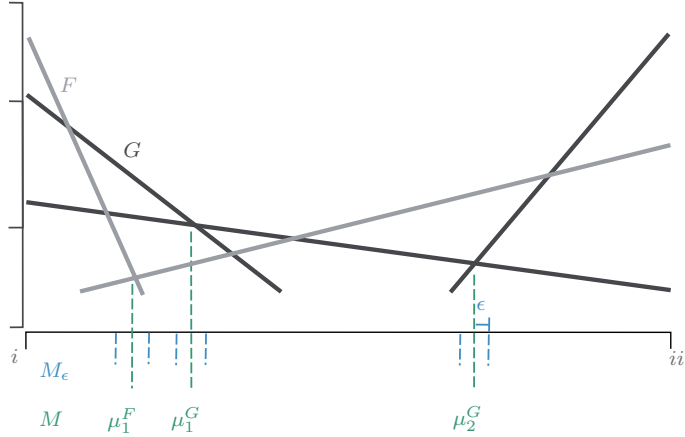
    }
    \caption{Construction of the sets $M$ and $M_\epsilon$ for an example two follower game. The expected utilities of the follower $F$ (grey, 2 actions) and $G$ (black, 3 actions) are drawn as solid lines. The points of tie-breaking are marked in green and collected in the set $M$. The set $M_\epsilon$ (blue) includes points $\epsilon$-far from the points in $M$.}
    \label{fig:the set N}
\end{figure}

\begin{algorithm}
\caption{Approximately Optimal Ambiguous Strategy in Coupled $2 \times m$-Stackelberg Game.}
\label{alg:optimal ambiguous Stackelberg strategy}
\KwData{Leader utility matrices $U_{L(F)} \in \reals^{2\times m}$, follower utility matrices $U_F \in \reals^{2\times m}$ for all follower $F \in \cF$, follower tie-breaking rule $s$, approximation parameter $\epsilon > 0$}
\KwResult{$\epsilon$-approximate optimal ambiguous Stackelberg strategy $[\underline{p}^*, \overline{p}^*]$.}
(1) \For{each follower $F \in \cF$}{
    (i) Remove all weakly dominated strategies in the zero-sum game defined by $U_F$ \;
    (ii) Compute the best response landscape of the follower in the zero-sum game defined by $U_F$, \ie, $\mu^F_{1}, \ldots \mu^F_{n-1}$\;
}

(2) \begin{align*}
    M &\coloneqq \bigcup_{F \in \cF} \{0, \mu^F_{1}, \ldots \mu^F_{n-1}, 1\}\\
    M_\epsilon &\coloneqq \bigcup_{F \in \cF} \{\mu^F_{1} - \epsilon, \ldots \mu^F_{n-1} -\epsilon, 1-\epsilon\}  \cup \bigcup_{F \in \cF} \{\epsilon, \mu^F_{1} + \epsilon, \ldots \mu^F_{n-1} +\epsilon\}\\
    N &\coloneqq (M\cup M_\epsilon) \cap [0,1]
    \end{align*}\;
(3) Check whether $\epsilon < \frac{1}{2} \min_{\mu, \mu' \in M, \mu \neq \mu'} \left|\mu - \mu' \right|$, otherwise choose smaller $\epsilon$\;
(4) For all combinations $(\ell, u) \in N^2$, compute the ambiguous Stackelberg value $W([\ell, u])$ \;
(5) Output $[\underline{p}^*, \overline{p}^*] = \argmax_{[\ell, u] \colon (\ell, u) \in N^2} W([\ell, u])$ with the highest ambiguous Stackelberg value \;
\end{algorithm}
With the structure of the maxmin responses at our hand, we introduce Algorithm~\ref{alg:optimal ambiguous Stackelberg strategy}. The algorithm exploits the maxmin response landscape described in Lemma~\ref{lemma:maxmin best responses -- empty equal action} and Lemma~\ref{lemma:maxmin best responses -- non-empty equal action}. The search space of potential ambiguous commitments is reduced to intervals with endpoints in a set $N$. The set $N$ is constructed by two subsets $M$ and $M_\epsilon$. The set $M$ contains, for all followers $F \in \cF$, the points of tie-breaking in which the follower moves to respond with one action to responding with another action. If we would only use points in $M$ as endpoints of the committed intervals, then we potentially run into troubles with unfavorable tie-breaking. For certain configurations an interval with an endpoint $\epsilon$-far from a point in $M$ can invoke different responses than an interval with endpoint in $M$. Hence, we include $M_\epsilon$, the set of all points close to points in $M$. An exhaustive maximization search on all combinations of endpoints concludes the algorithm. For an illustration of the construction of $N$ see Figure~\ref{fig:the set N}.
\begin{remark} \label{rem:consistent}
    In the related work by \citet{tillio2016design} and \citet{dütting2024ambiguouscontracts} the authors assume that the leader only commits to consistent ambiguous strategies (\cf Section~\ref{sec:implementing ambiguity}). Perhaps surprisingly, Algorithm~\ref{alg:optimal ambiguous Stackelberg strategy} can be easily adapted to output an approximately optimal \emph{consistent} ambiguous Stackelberg strategy. To this end, one has to include an intermediate step in (4), which tests whether the interval $[\ell, u]$ is consistent. This is easily implemented by checking whether the expected payoff for the fixed responses of followers to $[\ell, u]$ under $\ell$ and $u$ is equal. 
\end{remark}

\begin{theorem}
\label{thm:poly-time algorithm for 2x m games}
    Let $\cG$ be a coupled $2 \times m$-Stackelberg game with followers $\cF$. None of the followers has a weakly dominated strategy in their payoff matrix. Let $C \coloneqq \max_{F \in \cF} \max_{a|F \in A_F} |U_{L_F}(ii, a_F) - U_{L_F}(i, a_F)|$. For any tie-breaking rule $s$ computing in $O(S(|\cF|, n))$ and a sufficienty small $\epsilon > 0$ the Algorithm~\ref{alg:optimal ambiguous Stackelberg strategy} provides an $\epsilon |\cF|C$-approximately optimal, ambiguous Stackelberg strategy in $O(|\cF|n^2 + n^2 |\cF|^2 S(|\cF|, n))$ runtime.
\end{theorem}
\begin{proof}
    Let us first prove the $\epsilon |\cF|C$-approximate optimality of the result. We introduce some notation. For an arbitrary interval $[v,w] \subseteq [0,1]$ the Stackelberg value is given by
    \begin{align*}
        W(s, [v,w]) &= \min_{p_L \in [v,w]} \sum_{F \in \mathcal{F}} U_{L_F}(p_L, s(\BR^a_F([v,w]))).
    \end{align*}
    We call $s(\BR^a_F([v,w])$ the maxmin response and the tuple $(b_F)_{F \in \cF}$ with $b_F \coloneq s(\BR^a_F([v,w]))$ the \emph{response pattern} to $[v,w]$. The maximal Stackelberg value achieved by Algorithm~\ref{alg:optimal ambiguous Stackelberg strategy} is given by
        \begin{align*}
         \hat{W}(s) \coloneqq \min_{p_L \in [\underline{p}^*, \overline{p}^*]} \left\langle p_L, \sum_{F \in \mathcal{F}} U_{L_F}(\cdot, s(\BR^a_F([\underline{p}^*, \overline{p}^*]))) \right\rangle = \max_{(\ell,u)\in N^2} \min_{p_L \in [\ell, u]} \left\langle p_L, \sum_{F \in \mathcal{F}} U_{L_F}(\cdot, s(\BR^a_F([\ell, u]))) \right\rangle.
    \end{align*}
    
    Note that for a fixed response pattern we can rewrite the sum in the Stackelberg values as a linear function,
    \begin{align*}
        \sum_{F \in \mathcal{F}}  U_{L_F}(p_L, b_F)
        = \sum_{F \in \mathcal{F}} \left\langle p_L, U_{L_F}(\cdot, b_F) \right\rangle
        = \left\langle p_L, \sum_{F \in \mathcal{F}} U_{L_F}(\cdot, b_F) \right\rangle,
    \end{align*}
    with slight abuse of notation in $p_L$ as mentioned in the introduction of this section. \emph{For the rest of the argument, we assume that the linear mapping is increasing in $p_L$. The argument analogously runs through in the decreasing case.}
    
    Crucial to the argument is the assumption that $\epsilon < \frac{1}{2} \min_{\mu, \mu' \in M, \mu \neq \mu'} \left|\mu - \mu' \right|$, since then for all pairs $\mu, \mu' \in M$, $\mu < \mu'$ implies $\mu + \epsilon < \mu'$ and $\mu > \mu'$ implies $\mu - \epsilon > \mu'$.

    To the end of showing $\epsilon |\cF| C$-approximate optimality, we argue that for any ambiguous commitment and its induced response pattern there is an ambiguous commitment which is an interval with endpoints in $N$ and achieves payoff at least as $\epsilon |\cF| C$-good as that original commitment. The argument is simple, but a tedious case study: roughly, for any given interval the closest points in $N$ are shown to induce the same response pattern as the interval, but yields an approximately higher worst-case expected payoff then the interval. We have moved the argument to the appendix (Proof~\ref{proof:polytime algorithm gives optimal commitment}). In summary, Proof~\ref{proof:polytime algorithm gives optimal commitment} implies that 
    \begin{align*}
        \hat{W}(s) \ge W^*(s) - \epsilon |\cF| C.
    \end{align*}
    Hence, Algorithm~\ref{alg:optimal ambiguous Stackelberg strategy} outputs an $\epsilon |\cF| C$-approximately optimal, ambiguous Stackelberg strategy.

    Second, the runtime is a straightforward approximation of the computational steps. First, for the computation of the best response landscape, and second 
    \begin{align*}
        \begin{pmatrix}
            3 n |\cF|\\
            2
        \end{pmatrix} = \frac{3n |\cF| (3n |\cF| - 1)}{2} = O(n^2 |\cF|^2).
    \end{align*}
    Finally, we have to include the computation time for the tie-breaking rule.
\end{proof}

% \rab{I claimed that the following is true:}
% \begin{corollary}
%     There exists a poly-time tie-breaking rule $s^*$ such that the Algorithm~\ref{alg:optimal ambiguous Stackelberg strategy} provides an optimal ambiguous Stackelberg strategy in $O(|\cF|n^2 + n^2 |\cF|^2 S(|\cF|, n))$ runtime.
% \end{corollary}
% \rab{But Im now convinced that my definition of a tie-breaking rule, see below it is \emph{not} doing the job. The problem is that it might be advantageous for the leader that the follower best responds with a mixture of some of the endpoints of the maxmin best response set (this depends on the payoffs of the leader). But in the tie-breaking rule below, this can not be the case. Hence, even in the 2x m game it is not clear to me how a single follower can compute the tie-breaking rule which maximizes the worst-case of the expected payoff of the leader.}
% \begin{proof}
%     Let us first define the tie-breaking rule $s^*$. For a follower whose maxmin best response set is not a singleton we define
%     \begin{align*}
%         s^*(\BR^a([\ell, u])) \coloneqq \BR^a([\ell - \epsilon, u + \epsilon]))
%     \end{align*}
%     for an $\epsilon > 0$ as small as possible such that $|\BR^a([\ell - \epsilon, u + \epsilon]))| = 1$.
% \end{proof}

\section{Hardness Results}
\label{sec:hardness}
In this section we show that, when considering coupled Stackelberg games where $n = \Omega(k)$, computing the optimal ambiguous Stackelberg strategy is NP-hard. We do this via a reduction to min vertex cover. In Appendix~\ref{thm:pure_hard}, we show that the problem remains NP-hard even if we restrict ourselves to finding the optimal \emph{pure} ambiguous set. 

\begin{theorem}
For all simple tie-breaking rules, there exists a coupled Stackelberg game $\cG$ such that the problem of determining the optimal ambiguous commitment is NP-hard.
% If followers break the ties lexicographically\footnote{To formalize this: when deciding between different maxmin distributions, they break the ties in favor if which distribution plays their  first action with highest probability, then their second action, and so on.}, the problem of determining the optimal ambiguous commitment is NP-hard.
\end{theorem}
\begin{proof}
Note that throughout this proof we use the fact that when multiple leader actions have the exact same follower utility vectors, the maxmin distribution of the follower can be computed by considering all equivalent leader actions as a single action.

Let us assume we have a black-box algorithm $\A$ that takes as input a coupled Stackelberg game and outputs the optimal ambiguous commitment. We will use this to solve min-vertex cover. Given a graph $G$ with $n_{v}$ vertices, we will solve min-vertex cover for each connected  component separately. Thus, we can assume that the input to $G$ is connected. The input to $\A$ is as follows:
\begin{description}
\item[Include Games] $n$ games of size $n_{v}$ x $2$ indexed $I_{1}...I_{n_{v}}$. The follower in game $I_{i}$ has utilities as follows:

 \begin{table}[h]
        \centering
        \begin{tabular}{c|c c }
            & $a$ & $b$\\ \hline
            $a_L = i$ & 1 & 1  \\
            $a_L \neq i$ & 1 & 2  \\
        \end{tabular}
  \end{table}

The leader's payoff matrix for each of these games is $n_{v}^{2}$ whenever the follower plays $a$, and $0$ whenever the follower plays $b$, regardless of the leader action.  

\item[Exclude Games] $n$ games of size $n_{v}$ x $2$ indexed $E_{1}...E_{n_{v}}$. The follower in game $E_{i}$ has utilities as follows:
 \begin{table}[h]
        \centering
        \begin{tabular}{c|c c}
            & $a$ & $b$\\ \hline
            $a_L = i $ & 0 &  1\\
            $a_L \neq i $ & 1 &  1\\
        \end{tabular}
  \end{table}

%(If it's not in at all, they break the ties towards $a$. If it's in at all, they break the ties towards $b$.)

The leader's payoff matrix for each of these games is $n_{v}^{2}$ whenever the follower plays $a$, and $0$ whenever the follower plays $b$, regardless of the leader action.  

Together, the Include and Exclude games encode the constraint that the selection of vertices cannot be fractional.

\item[Cover Games] $k = n_{v}$ games of size $n_{v}$ x $2$, each associated with a particular vertex. The payoff for the leader is the same in all games: regardless of the leader's action, if the follower plays $a$, the leader gets payoff $1$, and if the follower plays $b$, the leader gets payoff $0$. From the follower in game $C_i$'s perspective, if vertex $j$ is adjacent to vertex $i$ or $j = i$, $u_{i}(j,a) = 0$ and $u_{i}(j,b) = -1$. Otherwise, $u_{i}(j,a) = 0$ and $u_{i}(j,b) = 1$. 
 \begin{table}[h]
        \centering
        \begin{tabular}{c|c c}
            $C_i$ & $a$ & $b$\\ \hline
            $a_L = i$ & 0 &  -1\\
            $a_L \in adj(i)$ & 0 &  -1\\
            $a_L \neq i, a_{L \notin adj(i)}$ & 0 & 1
        \end{tabular}
  \end{table} These games will represent the constraint that, given that the output must be integral, it must also represent a vertex cover. 
\item[Minimization Game] One game of size $n_{v}$ x $n_{v}$. For both the leader and the follower, their payoffs in $M$ are the identity matrix. This game encodes the constraint that, given that the output must represent a vertex cover, it must be a min vertex cover. 
\end{description}

We will prove that the output of $\A$ on this input, when re-interpreted as a set of vertices, is the min vertex cover. To see this, we will first prove a series of intermediary results:

\begin{lemma}In game $I_{i}$, the follower will play action $a$ if and only if the leader action $i$ is an extreme point of the set. Otherwise, he will play action $b$.  \label{lem:include}
\end{lemma}

\begin{proof}
If the Leader action $i$ is an extreme point of the set $S$, then $i$ is a dominant strategy for the Leader (regardless of what other extreme points there are in $S$). Thus, the follower's maxmin strategy is any best response to $i$. Both $a$ and $b$ are best responses to $i$, but since we are assuming simple tie-breaking, the follower will choose to play the pure strategy $a$, an extreme point of the set of responses. In case the simple tie-breaking rule selects the other extreme point, \ie, pure strategy $b$ is chosen, we simply switch the columns of the Include Games.
If the Leader action $i$ is not an extreme point of the set, then every extreme point of $S$ must give strictly more payoff to the follower when he plays $b$ over $a$. Thus $b$ is a strictly dominant strategy, so the follower will play pure strategy $b$. 
\end{proof}

\begin{lemma}
In game $E_{i}$, the follower will play action $a$ if and only if the leader action $i$ is not included in any mixed strategy in $S$. Otherwise, she will play action $b$.  \label{lem:exclude}
\end{lemma}

\begin{proof}
If any extreme point in $S$ puts support on Leader action $i$, then if the follower plays some distribution $p(a),p(b)$, the leader can play whatever distribution in $S$ puts most support on action $i$ and give the follower payoff strictly below $1$. If the follower plays action $b$, he is always guaranteed payoff $1$. Therefore the follower will play the pure strategy $b$. 
If no point in $S$ puts any support on action $i$, the entire game matrix is equal between $a$ and $b$. Since we are assuming simple tie-breaking, the follower will choose to play pure strategy $a$, an extreme point of the set of responses. In case the simple tie-breaking rule selects the other extrme point, \ie, pure strategy $b$ is chosen, we simply switch the columns of the Exclude Games.
\end{proof}

These two Lemmas immediately lead to the following corollary:

\begin{corollary}
  For every $i \in n_{v}$, at least one of the followers in games $I_{i}$ and $E_{i}$ will play the pure strategy action $b$.   \label{cor:IE}
\end{corollary}

Now, we are ready to describe the structure of the optimal ambiguous set output by $\A$. 

\begin{lemma}The output of $\A$ must be a set of pure strategies. \label{lem:pure}
\end{lemma}

\begin{proof}
Assume for contradiction that this is not the case. Then given some input graph $G$, the $S$ output by $\A$ has a pure strategy $t$ which: 1) is included in the mixed strategy of some extreme point in $S$, and 2) is not itself an extreme point of $S$. Then, by Lemmas~\ref{lem:include} and ~\ref{lem:exclude}, the followers in games $E_{i}$ and $I_{i}$ will both play action $b$. Keeping this and Corollary~\ref{cor:IE} in mind, we can write the payoff of the Leader as 

\begin{align*}
& \min_{s \in S} \left[\sum_{\forall I}u_{L,I_{i}}(s,BR^{a}_{F,I_{i}}(S)) + \sum_{\forall E}u_{L,E_{i}}(s,BR^{a}_{F,E_{i}}(S)) + \sum_{\forall C}u_{L,C_{i}}(s,BR^{a}_{F,C_{i}}(S)) + u_{L,M}(s,BR^{a}_{F,M}(S)) \right] \\
& \leq \min_{s \in S} \left[\sum_{\forall I}u_{L,I_{i}}(s,BR^{a}_{F,I_{i}}(S)) + \sum_{\forall E}u_{L,E_{i}}(s,BR^{a}_{F,E_{i}}(S))\right] + n_{v} + 1 \tag{By upper bounding the payoff in games $C$ and $M$ by the max-value action pair in the Leader's payoff matrix} \\
& = \sum_{\forall I}u_{L,I_{i}}(1,BR^{a}_{F,I_{i}}(S)) + \sum_{\forall E}u_{L,E_{i}}(1,BR^{a}_{F,E_{i}}(S)) + n_{v} + 1 \tag{As the leader's payoff is now agnostic to her realized action} \\
& \leq n_{v}^{2}\cdot (n_{v} - 1)  + n_{v} + 1  \tag{By the payoff matrices, Corollary~\ref{cor:IE} and the fact that both $E_{t}$ or $I_{t}$ play action $b$} \\
& = n_{v}^{3} - n_{v}^{2} + n_{v} + 1 
\end{align*}

Now, consider the following strategy: include every vertex. Then, the payoff of the Leader is

\begin{align*}
& \min_{s \in S} \left[\sum_{\forall I}u_{L,I_{i}}(s,BR^{a}_{F,I_{i}}(S)) + \sum_{\forall E}u_{L,E_{i}}(s,BR^{a}_{F,E_{i}}(S)) + \sum_{\forall C}u_{L,C_{i}}(s,BR^{a}_{F,C_{i}}(S)) + u_{L,M}(s,BR^{a}_{F,M}(S)) \right] \\
& \geq \min_{s \in S} \left[\sum_{\forall I}u_{L,I_{i}}(s,BR^{a}_{F,I_{i}}(S)) + \sum_{\forall E}u_{L,E_{i}}(s,BR^{a}_{F,E_{i}}(S))\right] \\
& = \sum_{\forall I}u_{L,I_{i}}(1,BR^{a}_{F,I_{i}}(S)) + \sum_{\forall E}u_{L,E_{i}}(1,BR^{a}_{F,E_{i}}(S))  \tag{As the leader's payoff is now agnostic to her realized action} \\
& \geq n_{v}^{2}\cdot n_{v}  = n_{v}^{3} \tag{By the fact that every follower in $I$ will play $a$}
\end{align*}

For large enough values of $n_{v}$, $n_{v}^{3} >  n_{v}^{3} - n_{v}^{2} + n_{v} + 1$. Therefore, our original set $S$ cannot be optimal, so we have derived a contradiction.

\end{proof}

We will now characterize the payoffs possible in the subset of the game matrices representing the Minimization games.

\begin{lemma}
\label{lemma: minimization game payoff}
    If $S$ is composed only of pure strategies, the leader's payoff in the minimization game $M$ is $\frac{1}{|S|}$.
\end{lemma}
\begin{proof}
    Consider any pure ambiguous strategy $S \subseteq [n_v]$. The maxmin best response of follower to $S$ is the uniform distribution on $S$ for two reasons. Any weight on an action $i \notin S$ cannot increase the maxmin utility for follower since the utilities are $0$. Any non-uniform  distribution on the actions $S$ leads to at least one action $i \in S$ with minimal weight $w_i < \frac{1}{|S|}$. In the worst case, this action is picked from $S$, hence the maxmin utility is $w_i$, which is smaller than the maxmin utility for the uniform distribution $\frac{1}{|S|}$.
    Finally, it follows from the utility matrix of the leader that the worst-case expected utility given the uniform best response is $\frac{1}{|S|}$ for the leader, too.
\end{proof}

Next, we can combine these ideas together to show that, if the agents each play action $a$, the strategies in $S$ correspond to a vertex cover.

\begin{lemma} \label{lem:vc}
If $S$ is composed only of pure strategies, the maxmin strategies of the followers in the Cover games are always pure strategies, and the output of $\cA$ corresponds to a vertex cover.
\end{lemma}
\begin{proof}
To see this, note that if some vertex $i$ is not covered, the utility matrix for the follower in game $C_i$ includes only actions that strictly incentivize action $b$. Therefore, he will play pure strategy $b$. On the other hand, if the vertex is covered, the follower's utility matrix looks as follows:

 \begin{table}[h]
        \centering
        \begin{tabular}{c|c c}
             & $a$ & $b$\\ \hline
            $a_L = i$ & 0 &  -1\\
            $a_L \in adj(i)$ & 0 &  -1\\
            $a_L \neq i, a_{L \notin adj(i)}$ & 0 & 1
        \end{tabular}
  \end{table}

Given some distribution $p(a), p(b)$ of the follower, his payoff against a leader playing action $i$ is $-p(b)$, while his payoff against a leader playing any other action is $p(b)$. Therefore the worst-case for any follower distribution is attained against action $i$, so the follower's maxmin is the best response to action $i$. This is the pure strategy $a$. Therefore, each follower will either play $a$ or $b$, and he will play $a$ iff his corresponding vertex is covered in the pure ambiguous set. 

Next, we will show that the output to $\A$ must be a vertex cover. Assume for contradiction that this is not the case. Then we have some optimal ambiguous set $S$ which, from above, has at least one follower $i$ play action $b$. The payoff of the leader is equal to the number of followers that play action $a$ in the Cover games, plus $\frac{1}{|S|}$ (from the Minimization games; Lemma~\ref{lemma: minimization game payoff} ), plus at most $n_{v}^{2} \cdot n$ (the maximum value that can be jointly attained from the Include and Exclude games). Therefore, the payoff of $S$ is at most $n_{v}^{3} + (n_{v} - 1) + \frac{1}{|S|} \leq n_{v}^{3} + (n_{v} - 1) + 1 = n_{v}^{3} + n_{v}$. Now, consider the payoff of full ambiguity. Every follower is covered, and $\A$ is a set of pure strategies, so the payoff is $n^{3}_{v} + n_{v} + \frac{1}{n_{v}}$. This is larger, and therefore $S$ cannot be optimal. This proves a contradiction, and therefore if the output to $\A$ is integral, it must be a vertex cover.
\end{proof}

%\begin{lemma}The pure strategies output by $\A$ must correspond to a vertex cover in $G$, where $S_{i}$ corresponds to $V_{i}$.
%\end{lemma}
Now we can put everything together:

\begin{lemma}The output of $\A$ must correspond to a min-vertex cover in $G$.
\end{lemma}
\begin{proof}
By Lemma~\ref{lem:pure}, the output of $\A$ is composed only of pure strategies. Therefore by Lemma~\ref{lem:vc}, the output of $\A$ corresponds to a vertex cover. 

Note that the utility for the leader for every vertex cover is $n_{v}^{3} + n_{v} + \frac{1}{|S|}$ (Lemma~\ref{lemma: minimization game payoff}). This is maximized when $|S|$ is minimized. Therefore, for $\A$ to be optimal, the solution output by $\A$ must be not only a vertex cover but a min-vertex cover. 
\end{proof}

\end{proof}

\bibliography{main}
\bibliographystyle{plainnat}

\section{Appendix}

\subsection{Notes on Zero Sum Games}
%\rab{Can we find a similar lemma as the one below where the follower does not make all actions of the leader invariant, but only a subset, and then playing full ambiguity on the subset is the best one can do?}

The statement below is related to \emph{immunization strategies} in [Ellsberg Games 2013, Definition 2].

\begin{lemma}
\label{lemma:invariance to actions - full ambiguous strategy}
Let $\cG$ be a coupled, zero-sum Stackelberg game with $u_F = -u_{L_F}$ for all $F \in \cF$.
If the maxmin-strategies $\BR^{a}_F(\Delta(A_L))$ of all followers $F \in \cF$ make the leader invariant to her actions, \ie
for $b_F \in \BR^{a}_F(\Delta(A_L))$
\begin{align}
\label{eq:invariance of actions for leader given maxmin strategy of follower}
    U_F(a_L, b_F) = U_F(a_L', b_F), 
\end{align}
for all pairs of actions for leader $a_L, a_L' \in A_L$,
then the leader cannot do better than using full ambiguity, independent of the chosen tie-breaking rule $s$.
\end{lemma}
\begin{proof}
For any $P_L \in \cA(A_L)$, we have that (for any tie-breaking rule $s$),
\begin{align*}
    \min_{p_L \in P_L} U_F(p_L, s(\BR^{a}(P_L))) &= \max_{p_F \in \Delta(A_F)} \min_{p_L \in P_L} U_F(p_L, p_F)\\
    &\ge \max_{p_{F} \in \Delta(A_F)} \min_{p_L \in \Delta(A_L)} U_F(p_F, p_L)\\
    &\eqqcolon maxmin_F.
\end{align*}

For a fixed $P_{L}$, we can write the payoff of the leader. Recall that the leader is a minimizer and wants the value to be as low as possible.
\begin{align*}
 W(P_L, s) &\coloneqq \min_{p_L \in P_L}\sum_{F \in \cF} U_{L_F}(p_L, s(\BR^{a}_F(P_L))) \\
 &= -\max_{p_L \in P_L}\sum_{F \in \cF} U_F(p_L, s(\BR^{a}_F(P_L))) \\
 & \leq -\min_{p_L \in P_L} \sum_{F \in \cF} U_F(p_L, s(\BR^{a}_F(P_L))) \\
& \leq -\sum_{F \in \cF} \min_{p_L \in P_L}  U_F(p_L, s(\BR^{a}_F(P_L))) \\
& \leq -\sum_{F \in \cF} maxmin_{F}.
\end{align*}
Thus, regardless of the choice of $P_L$, the payoff of the game cannot be more than $-\sum_{F \in \cF} maxmin_{F}$. 

Finally, we will show that picking $P_L = \Delta(A_L)$ makes the payoff of the game exactly $- \sum_{F \in \cF} maxmin_{F}$.
To see this, note that (for any tie-breaking rule $s$)
\begin{align*}
    W(\Delta(A_L), s) &= \min_{p_L \in P_L}\sum_{F \in \cF} -U_{L_F}(p_L, s(\BR^{a}_F(\Delta(A_L))))\\
    &= -\max_{p_L \in P_L}\sum_{F \in \cF} U_F(p_L, s(\BR^{a}_F(\Delta(A_L))))\\
    &\ge - \sum_{F \in \cF} \max_{p_L \in P_L} U_F(p_L, s(\BR^{a}_F(\Delta(A_L))))\\
    &\overset{\eqref{eq:invariance of actions for leader given maxmin strategy of follower}}{=} -\sum_{F \in \cF} \min_{p_L \in P_L} U_F(p_L, s(\BR^{a}_F(\Delta(A_L))))\\
    &= -\sum_{F \in \cF} maxmin_{F},
\end{align*}
concluding the proof.
\end{proof}

The assumption that the leader is invariant in her actions given the follower's maxmin strategy \eqref{eq:invariance of actions for leader given maxmin strategy of follower} is rather strict. In games where there exists a (weakly) dominated strategy for leader this is never case.

\begin{corollary}
    Let $\cG$ be a coupled, zero-sum $2 \times m$-Stackelberg game and leader has no dominated action in any sub-game $\cG_F$ for all $F \in \cF$ of the corresponding decoupled, zero-sum Stackelberg game family $(\cG_F)_{F \in \cF}$. Then leader cannot do better than using full ambiguity.
\end{corollary}
\begin{proof}
    In the case that leader has no dominated action in any sub-game for all $F \in \cF$, each follower ties the two possible actions of the leader. This way follower can achieve the maxmin value in the sub-game. Additionally, it implies that it makes the leader invariant with respect to her actions. Hence, Lemma~\ref{lemma:invariance to actions - full ambiguous strategy} applies.
\end{proof}

\subsection{Bounds on the Individualized Stackelberg Value}

\begin{lemma} If the utilities of the leader are always non-negative, then \begin{equation}
   V^{*}(s) \geq \frac{ISV}{k}
\end{equation} for some tie-breaking rule $s$. \label{lem:ISV_UB}
\end{lemma}
\begin{proof}

Let $p^{*} = \argmax_{p_{L} \in \Delta(A_{L})}\sum_{F \in \mathcal{F}}U_{L_{F}}(p_{L},s(BR_{F}(p_{L}))$. Furthermore, let $\bar{F} = \argmax_{F \in \mathcal{F}}V^{*}_{F}(\bar{s})$, and let $\bar{p}$ be the classic Stackelberg distribution of the leader against follower $\bar{F}$. Then we have,
\begin{align*}
    ISV & = \sum_{F \in \mathcal{F}}V^{*}_{F}(\bar{s}) \\
    & \leq k \cdot \max_{F \in \mathcal{F}}V^{*}_{F}(\bar{s}) \\
    & = k \cdot V^{*}_{\bar{F}}(\bar{s}) \tag{By the definition of $\bar{F}$}
    \\ & = k \cdot U_{L_{\bar{F}}}(\bar{p},\bar{s}(BR_{\bar{F}}(\bar{p}))) \tag{By the definition of $\bar{p}$}
    \\ & \leq k \cdot \sum_{F \in \mathcal{F}}U_{L_{F}}(\bar{p},\bar{s}(BR_{F}(\bar{p}))) \tag{By fact that all values of $U_{L_{F}}$ are non-negative}
    \\ & \leq k \cdot \max_{p_{L} \in \Delta(\A_{L})} \sum_{F \in \mathcal{F}}U_{L_{F}}(p_{L},\bar{s}(BR_{F}(p_{L})))
    \\ & = k \cdot V^{*}(s). \tag{For the tie-breaking rule $s$ where each follower maximizes the leader's utility on his own sub-game} 
\end{align*}

\end{proof}

\begin{lemma}  \label{lem:ISV_lb} \begin{equation}
   V^{*}(s) \leq ISV
\end{equation}
\end{lemma}
\begin{proof}
\begin{align*}
    V^{*}(s) & = \max_{p_{L} \in \Delta(\A_{L})} \sum_{F \in \mathcal{F}}U_{L_{F}}(p_{L},s(BR_{F}(p_{L}))) \\
    & \leq \sum_{F \in \mathcal{F}}\max_{p_{L} \in \Delta(\A_{L})}  U_{L_{F}}(p_{L},s(BR_{F}(p_{L}))) \\
    & \leq \sum_{F \in \mathcal{F}}\max_{p_{L} \in \Delta(\A_{L})}  U_{L_{F}}(p_{L},\bar{s}(BR_{F}(p_{L}))) \tag{Where $\bar{s}$ is the tie-breaking rule which maximizes in favor of the leader in each individual game} \\
    & = ISV \\
\end{align*}

\end{proof}

\begin{lemma} If all of the leader's utilities are positive in some coupled, zero-sum game, then $G(s) \leq k$. \label{lem:positive_ub}
\end{lemma}
\begin{proof}
Let $p^{*} = \argmax_{p_{L} \in \Delta(A_{L})}\sum_{F \in \mathcal{F}}U_{L_{F}}(p_{L},s(BR_{F}(p_{L}))$. Then, we have:

    \begin{align*}
        G(s) & = \frac{W^{*}(s)}{V^{*}(s)} \tag{As $W^{*}(s)$ and $V^{*}(s)$ are strictly positive} \\
        & \leq \frac{W^{*}(s)k}{ISV} \tag{By Lemma~\ref{lem:ISV_UB}, as all leader utilities are non-negative} 
        \\ & = k \cdot \frac{W^{*}(s)}{ISV} = k \cdot C(s) \\
        & \leq k \tag{By Theorem~\ref{thm:Bounds to Ambiguous Stackelberg Value}}
    \end{align*}
\end{proof}

\subsection{Additional Proofs from Section~\ref{sec:poly-time}}

\begin{lemma}
\label{lemma: Best response sets are closed and convex}
    The best response set for mixed strategies is closed and convex, \ie, let $p,q \in \Delta(A_L)$ such that $a \in \BR(q)$ and $a \in \BR(p)$ and $\alpha \in [0,1]$ then $a \in \BR(\alpha q + (1-\alpha)p)$ and for a sequence $(p_i)_{i \in \mathbb{N}}$ with limit point $\lim_{i\rightarrow \infty} p_i = p$ and $a \in \BR(p_i)$ it holds $a \in \BR(p)$.
\end{lemma}
\begin{proof}
    Note that
    \begin{align*}
        \max_{a_F \in A_F} \mathbb{E}_{a_L \sim \alpha q + (1-\alpha)p}[u_F(a_L, a_F)] &= \max_{a_F \in A_F} \alpha \mathbb{E}_{a_L \sim q}[u_F(a_L, a_F)] + (1-\alpha) \mathbb{E}_{a_L \sim p}[u_F(a_L, a_F)]\\
        &\le \max_{a_F \in A_F} \alpha \mathbb{E}_{a_L \sim q}[u_F(a_L, a_F)] + \max_{a_F \in A_F} (1-\alpha) \mathbb{E}_{a_L \sim p}[u_F(a_L, a_F)]\\
        &= \alpha \mathbb{E}_{a_L \sim q}[u_F(a_L, a)] + (1-\alpha) \mathbb{E}_{a_L \sim p}[u_F(a_L, a)]\\
        &=\mathbb{E}_{a_L \sim \alpha q + (1-\alpha)p}[u_F(a_L, a)]\\
        &\le \max_{a_F \in A_F} \mathbb{E}_{a_L \sim \alpha q + (1-\alpha)p}[u_F(a_L, a_F)].
    \end{align*}
    by Jensen's inequality and $a \in \BR(p)$ and $a \in \BR(q)$. Hence, $a \in \BR(\alpha q + (1-\alpha)p)$.

    For the closure property note that $\mathbb{E}_{a_L \sim p}[u_F(a_L, a_F)]$ can be rewritten as linear function $\langle p, u_F(\cdot, a_F)\rangle$, hence
    \begin{align*}
        \max_{a_F \in A_F} \mathbb{E}_{a_L \sim p_i}[u_F(a_L, a_F)] &= \max_{a_F \in A_F} \langle p, u_F(\cdot, a_F)\rangle\\
        &= \max_{a_F \in A_F} \langle \lim_{i \rightarrow \infty} p_i, u_F(\cdot, a_F)\rangle\\
        &= \max_{a_F \in A_F} \lim_{i \rightarrow \infty} \langle p_i, u_F(\cdot, a_F)\rangle\\
        &= \lim_{i \rightarrow \infty} \max_{a_F \in A_F}  \langle p_i, u_F(\cdot, a_F)\rangle\\
        &= \lim_{i \rightarrow \infty} \langle p_i, u_F(\cdot, a)\rangle\\
        &= \langle p, u_F(\cdot, a)\rangle,
    \end{align*}
    which concludes the proof.
\end{proof}

\begin{lemma}[Support of Maxmin Best Response]
\label{lemma:support of maxmin best response}
    Let $\cG$ be a coupled Stackelberg game with a single follower. Assume there are no weakly dominated actions for follower. Let $P_L \in \cA(A_L)$ be an ambiguous commitment. Define $B \coloneqq \{ a \in A_F\colon \BR^{-1}(a) \cap P \neq \emptyset\}$. Then, $\BR^a(P_L) \subseteq \Delta(B)$.
\end{lemma}
\begin{proof}
    \begin{align*}
        \max_{p_F \in \Delta(A_F)} \min_{p_L \in P_L} \mathbb{E}_{a_F \sim p_F, a_L \sim p_L}[u_F(a_L, a_F)] &\overset{(a)}{=} \min_{p_L \in P_L} \max_{p_F \in \Delta(A_F)}  \mathbb{E}_{a_F \sim p_F, a_L \sim p_L}[u_F(a_L, a_F)]\\
        &\overset{(b)}{=} \min_{p_L \in P_L} \max_{a \in B}  \mathbb{E}_{a_L \sim p_L}[u_F(a_L, a)]\\
        &= \min_{p_L \in P_L} \max_{p_F \in \Delta(B)}  \mathbb{E}_{a_F \sim p_F, a_L \sim p_L}[u_F(a_L, a)]\\
        &\overset{(c)}{=} \max_{p_F \in \Delta(B)}\min_{p_L \in P_L}  \mathbb{E}_{a_F \sim p_F, a_L \sim p_L}[u_F(a_L, a)].
    \end{align*}

    \begin{enumerate}[(a)]
        \item Because $P_L$ and $\Delta(A_F)$ are compact (closed and bounded) and convex and obviously the expectation operator is linear in both arguments, hence von Neumann's minmax theorem applies.
        \item Since $\BR(p_L) \in B$ for all $p_L \in P_L$.
        \item Because $P_L$ and $\Delta(B)$ are compact (closed and bounded) and convex and obviously the expectation operator is linear in both arguments, hence von Neumann's minmax theorem applies.
    \end{enumerate}
\end{proof}

\begin{lemma}[Tying the Leader]
\label{lemma: tying the leader}
    Let $\cG$ be a coupled $2 \times m$-Stackelberg game with a single follower. Let the follower have no weakly dominated strategy in $u_F$. Assume that $A^= = \emptyset$. Let
    \begin{align*}
         \nu \coloneqq \frac{1}{\frac{v_y-w_y}{w_x-v_x} + 1}.
    \end{align*}
    We define the action $t_p \coloneqq (1-p)\cdot a_x + p \cdot a_y$.
    For any choice of ambiguous commitment $[\underline{p}, \overline{p}] \subseteq [0,1]$ such that $\underline{p} < \overline{p}$, the action $t_\nu$ ties the actions $\underline{p}$ and $\overline{p}$, \ie,
    \begin{align*}
        U_F(\underline{p}, t_\nu) = U_F(\overline{p}, t_\nu).
    \end{align*}
    In particular, for $p < \nu$ (if $p \in [0,1]$),
    \begin{align*}
        U_F(\underline{p}, t_p) > U_F(\overline{p}, t_p).
    \end{align*}
    The analogous result holds for $p > \nu$ (if $p \in [0,1]$),
    \begin{align*}
        U_F(\underline{p}, t_p) < U_F(\overline{p}, t_p).
    \end{align*}
\end{lemma}
\begin{proof}
    First, note that (a) $w_x - v_x \neq 0$ because $A^= = \emptyset$, (b) $w_y - v_y \neq 0$ because $A^= = \emptyset$, (c) $v_x < w_x$ since $U_x$ has negative slope, and (d) $v_y > w_y$ since $U_y$ has positive slope, hence $\frac{v_y-w_y}{w_x-v_x} \in (0, \infty)$. This implies $\nu \in (0,1)$.
    
    We define
    \begin{align*}
        \begin{pmatrix}
            a & b \\
            c & d
        \end{pmatrix} \coloneqq
        \begin{pmatrix}
            (1-\underline{p}) w_x + \underline{p}v_x & (1-\underline{p}) w_y + \underline{p}v_y\\
            (1-\overline{p}) w_x + \overline{p}v_x & (1-\overline{p}) w_y + \overline{p}v_y
        \end{pmatrix}.
    \end{align*}
    Let us rewrite,
    \begin{align*}
        U_F(\underline{p}, t_p) &= a (1-p) + b p,\\
        U_F(\overline{p}, t_p) &= c (1-p) + d p.
    \end{align*}
    Note that
    \begin{align*}
        d-b &= ((1-\overline{p}) w_y + \overline{p}v_y) - ((1-\underline{p}) w_y + \underline{p}v_y)\\
        &= ((1- \overline{p}) - (1- \underline{p})) w_y + (\overline{p} - \underline{p}) v_y\\
        &= \Delta v_y - \Delta w_y,
    \end{align*}
    and
    \begin{align*}
        a - c &= ((1- \underline{p}) w_x + \underline{p} v_x) - ((1- \overline{p}) w_x + \overline{p} v_x)\\
        &= ((1-\underline{p}) - (1-\overline{p})) w_x + (\underline{p} - \overline{p}) v_x\\
        &= \Delta w_x - \Delta v_x.
    \end{align*}
    
    Hence,
    \begin{align*}
        & &U_F(\underline{p}, t_\nu) &= U_F(\overline{p}, t_\nu)\\
        &\Longleftrightarrow &a (1-\nu) + b \nu &= c (1-\nu) + d \nu\\
        &\Longleftrightarrow &(a-c) (1-\nu) &= (d-b) \nu\\
        &\overset{\nu \neq 0}{\Longleftrightarrow} &\frac{1-\nu}{\nu} &= \frac{d-b}{a-c}\\
        &\Longleftrightarrow &\frac{1}{\nu} &= \frac{d-b}{a-c} + 1\\
        &\overset{\frac{v_y-w_y}{w_x-v_x} \neq -1}{\Longleftrightarrow} &\nu &= \frac{1}{\frac{d-b}{a-c} + 1}\\
        &\Longleftrightarrow &\nu &= \frac{1}{\frac{v_y-w_y}{w_x-v_x} + 1},
    \end{align*}
    indifferent to the size and place of the ambiguous commitment $[\overline{p}, \underline{p}]$.

    In particular,
    \begin{align*}
        & &U_F(\underline{p}, t_p) &> U_F(\overline{p}, t_p)\\
        &\Longleftrightarrow &a (1-p) + b p &> c (1-p) + d p\\
        &\Longleftrightarrow &(a-c) (1-p) &> (d-b) p\\
        &\overset{p \neq 0}{\Longleftrightarrow} &\frac{1-p}{p} &> \frac{d-b}{a-c}\\
        &\Longleftrightarrow &\frac{1}{p} &> \frac{d-b}{a-c} + 1\\
        &\overset{\frac{v_y-w_y}{w_x-v_x} \neq -1}{\Longleftrightarrow} &p &< \frac{1}{\frac{d-b}{a-c} + 1}\\
        &\Longleftrightarrow &p &< \nu = \frac{1}{\frac{v_y-w_y}{w_x-v_x} + 1},
    \end{align*}
    indifferent to the size and place of the ambiguous commitment $[\overline{p}, \underline{p}]$. The analogous equivalences hold for $<$.
\end{proof}

\begin{lemma}[Lemma~\ref{lemma:maxmin best responses -- empty equal action}]
    Let $\cG$ be a coupled $2 \times m$-Stackelberg game with a single follower. Let the follower have no weakly dominated strategy in $u_F$. Assume that $A^= =\emptyset$. Define the point of slope sign change as $\mu^\pm$. The action left of the slope sign change is $a_x$, the action right of it is $a_y$.
    Let
    \begin{align*}
        \nu\coloneqq \frac{1}{\frac{v_y - w_y}{w_x - v_x} + 1}.
    \end{align*}
    We define the action $t_p \coloneqq (1-p)\cdot a_x + p \cdot a_y$
    \begin{enumerate}[(a)]
        \item If $ \underline{p} \le \overline{p} < \mu^\pm$, then $\BR^a(\upop) = \BR(\overline{p})$.
        \item If $\underline{p} < \overline{p} = \mu^\pm$, then $\BR^a(\upop) = \Delta(\{a_x, t_\nu\}) = \{t_p \colon p \in [0, \nu] \}$.
        \item If $\underline{p} = \overline{p} = \mu^\pm$, then $\BR^a(\upop) = \Delta(\{a_x,a_y\}) = \BR(\mu^\pm)= \{t_p \colon p \in [0, 1] \}$.
        \item If $ \mu^\pm = \underline{p} \le \overline{p}$, then $\BR^a(\upop) = \Delta(\{t_\nu, a_y\}) = \{t_p \colon p \in [\nu, 1] \}$.
        \item If $\underline{p} <  \mu^\pm <  \overline{p}$, then $\BR^a(\upop) = \{ t_\nu\}$.
        \item If $\mu^\pm < \underline{p} \le \overline{p} $, then $\BR^a(\upop) = \BR(\underline{p})$.
    \end{enumerate}
\end{lemma}
\begin{proof}
\label{proof:maxmin best responses -- empty equal action}
    \begin{enumerate}[(a)]
        \item
        \begin{align*}
            \max_{p_F \in \Delta(A_F)}\min_{p_L \in \upop} U_F(p_L, p_F) &\overset{(i)}{=} \max_{p_F \in \Delta(A^-)}\min_{p_L \in \upop} U_F(p_L, p_F)\\
            &\overset{(ii)}{=} \min_{p_L \in \upop} \max_{p_F \in \Delta(A^-)} U_F(p_L, p_F)\\
            &\overset{(iii)}{=} \max_{p_F \in \Delta(A^-)} U_F(\overline{p}, p_F)\\
            &\overset{(iv)}{=} \max_{p_F \in \Delta(A_F)} U_F(\overline{p}, p_F)\\
            &= U_F(\overline{p}, p^*_F),
        \end{align*}
        for any $p^*_F \in \BR(\overline{p})$.
        \begin{enumerate}[(i)]
            \item Note that $\bigcup_{p \in \upop} \BR(p) \subseteq A^-$ (Lemma~\ref{lemma:best response landscape}). Lemma~\ref{lemma:support of maxmin best response} implies $\BR^a(\upop) \subseteq \Delta(A^-)$.
            \item Von Neumann's minmax theorem.
            \item The negative slope implies
            \begin{align*}
                U_a(\underline{p}) \ge U_a(\overline{p}),
            \end{align*}
            for all actions $a \in A^-$ and their convex combinations.
            \item $\BR(\overline{p}) \subseteq \Delta(A^{-})$.
        \end{enumerate}

        \item
        We first provide an argument that $\BR^a(\upop) \supseteq \Delta(\{a_x, t\})$, let $q_F \in \Delta(\{ a_x, t\})$
        \begin{align*}
            \min_{p_L \in \upop}  U_F(p_L, q_F) &\overset{(i)}{=} U_F(\overline{p}, q_F)\\
            &\overset{(ii)}{=} \max_{p_F \in \Delta(A_F)}  U_F(\overline{p}, p_F)\\
            &\ge \min_{p_L \in \upop} \max_{p_F \in \Delta(A_F)}  U_F(p_L, p_F)\\
            &\overset{(iii)}{=} \max_{p_F \in \Delta(A_F)}  \min_{p_L \in \upop} U_F(p_L, p_F).
        \end{align*}
        \begin{enumerate}[(i)]
            \item For $p_F = a_x$, because of the negative slope $\min_{p_L \in \upop}  U_F(p_L, a_x) =  U_F(\overline{p}, a_x)$. The follower's action $p_F = t$ ties all leader's actions $p \in \upop$ (Lemma~\ref{lemma: tying the leader}), hence $\min_{p_L \in \upop}  U_F(p_L, t) =  U_F(\overline{p}, t)$. All convex combinations fulfill the same requirement.
            \item Because $\max_{p_F \in \Delta(A_F)}  U_F(\overline{p}, p_F) =  U_F(\overline{p}, p^*_F)$ for any $p^*_F \in \BR(\overline{p})$ and $a_x, t \in \BR(\overline{p})$. The last statement is an implication of the fact that $\BR(\overline{p}) = \BR(\mu^{\pm}) \supseteq \{ a_x, a_y\}$.
            \item Von Neumann's minmax theorem.
        \end{enumerate}
        
        Now let us argue that there is no other action $q \in \Delta(A_F) \setminus \Delta(\{ a_x, t\})$ such that $q \in \BR^a(\upop)$: 
        since we assumed that there are no weakly dominated strategies for follower,
        \begin{align*}
            U_F(\overline{p}, q_F) = U(\mu^\pm, q_F) > U(\mu^\pm, a),
        \end{align*}
        for all $a \in A_F \setminus \{ a_x, a_y\}$. Hence, only convex combinations of $a_x$ and $a_y$ can achieve the maxmin-value above. Note that 
        $\Delta(\{ a_x, t\}) = \{ t_p \colon 0 \le p \le \nu\} \subseteq \Delta( \{ a_x, a_y\})$
        If $p > \nu$, then $t_p$ has $U_F(\underline{p}, t_p) < U_F(\overline{p}, t_p)$, hence $\min_{p_L \in \upop}  U_F(p_L, t_p) < U_F(\overline{p}, t_p) = U_F(\overline{p}, t_\nu)$.

        \item Follows from the standard best responses to precise probabilities.
        \item As above.
        \item
        We first provide an argument that $\BR^a(\upop) \supseteq \{t_\nu\}$,
        \begin{align*}
            \min_{p_L \in \upop}  U_F(p_L, t_\nu) &\overset{(i)}{=} U_F(\mu^\pm, t_\nu)\\
            &\overset{(ii)}{=} \max_{p_F \in \Delta(A_F)}  U_F(\mu^\pm, p_F)\\
            &\ge \min_{p_L \in \upop} \max_{p_F \in \Delta(A_F)}  U_F(p_L, p_F)\\
            &\overset{(iii)}{=} \max_{p_F \in \Delta(A_F)}  \min_{p_L \in \upop} U_F(p_L, p_F).
        \end{align*}
        \begin{enumerate}[(i)]
            \item The follower's action $p_F = t_\nu$ ties all leader's actions $p \in \upop$ (Lemma~\ref{lemma: tying the leader}), hence $\min_{p_L \in \upop}  U_F(p_L, t_\nu) =  U_F(\overline{p}, t_\nu)$.
            \item Because $\max_{p_F \in \Delta(A_F)}  U_F(\mu^{\pm}, p_F) =  U_F(\mu^{\pm}, p^*_F)$ for any $p^*_F \in \BR(\mu^{\pm})$ and $t_\nu \in \BR(\overline{p})$. This is an implication of the fact that $\BR(\mu^{\pm}) \supseteq \{ a_x, a_y\}$.
            \item Von Neumann's minmax theorem.
        \end{enumerate}
        Now let us argue that there is no other action $q \in \Delta(A_F)$ such that $q \in \BR^a(\upop)$: since we assumed that there are no weakly dominated strategies for follower,
        \begin{align*}
            U(\mu^\pm, t) > U(\mu^\pm, a),
        \end{align*}
        for all $a \in A_F \setminus \{ a_x, a_y\}$. Hence, only convex combinations of $a_x$ and $a_y$ can achieve the maxmin-value above. The only such convex combination is $t_\nu$, \ie, $\min_{p_L \in \upop}  U_F(p_L, t_q) = U_F(\mu^\pm, t_q)$ if and only if $q = \nu$ (Lemma~\ref{lemma: tying the leader}).
        \item As above.
    \end{enumerate}
\end{proof}

\begin{lemma}[Lemma~\ref{lemma:maxmin best responses -- non-empty equal action}]
    Let $\cG$ be a coupled $2 \times m$-Stackelberg game with a single follower. Let the follower have no weakly dominated strategy in $u_F$. We assume $\cA^= = \{ a_=\}$ (without loss of generality it is a singleton). We define $\mu^{-=}$ such that $U_{x}(\mu^{-=}) = U_{a_=}(\mu^{-=})$ and $\mu^{=+}$ such that $U_{y}(\mu^{=+}) = U_{a_=}(\mu^{=+})$. The action left of the slope sign change is $a_x$, the action right of it is $a_y$.
    \begin{enumerate}[(a)]
        \item If $ \underline{p} \le \overline{p} < \mu_{=+}$, then $\BR^a(\upop) = \BR(\overline{p})$.
        \item If $\mu_{-=} < \underline{p} \le \overline{p} $, then $\BR^a(\upop) = \BR(\underline{p})$.
        \item If $\underline{p} \le  \mu_{-=} < \mu_{=+} \le \overline{p}$, then $\BR^a(\upop) = \{a_=\}$.
    \end{enumerate}
\end{lemma}
\begin{proof}
\label{proof:maxmin best responses -- non-empty equal action}
    The set $A^=$ is, without loss of generality, a singleton. The reason for this is because otherwise the set contains two identical actions respectively one action strictly dominates all others in the set. But we assumed that follower has no weakly dominated strategy.
    \begin{enumerate}[(a)]
        \item \begin{align*}
            \max_{p_F \in \Delta(A_F)}\min_{p_L \in \upop} U_F(p_L, p_F) &\overset{(i)}{=} \max_{p_F \in \Delta(A^-)}\min_{p_L \in \upop} U_F(p_L, p_F)\\
            &\overset{(ii)}{=} \min_{p_L \in \upop} \max_{p_F \in \Delta(A^-)} U_F(p_L, p_F)\\
            &\overset{(iii)}{=} \max_{p_F \in \Delta(A^-)} U_F(\overline{p}, p_F)\\
            &\overset{(iv)}{=} \max_{p_F \in \Delta(A_F)} U_F(\overline{p}, p_F)\\
            &= U_F(\overline{p}, p^*_F),
        \end{align*}
        for any $p^*_F \in \BR(\overline{p})$.
        \begin{enumerate}[(i)]
            \item Note that $\bigcup_{p \in \upop} \BR(p) \subseteq A^-\cup A^=$ (Lemma~\ref{lemma:best response landscape}). Lemma~\ref{lemma:support of maxmin best response} implies $\BR^a(\upop) \subseteq \Delta(A^- \cup A^=)$.
            \item Von Neumann's minmax theorem.
            \item The non-positive slope implies
            \begin{align*}
                U_a(\underline{p}) \ge U_a(\overline{p}),
            \end{align*}
            for all actions $a \in A^- \cup A^=$ and their convex combinations.
            \item $\BR(\overline{p}) \subseteq \Delta(A^{-} \cup A^=)$.
        \end{enumerate}

        \item Analogous argument as above applies.
        \item We first provide an argument that $\BR^a(\upop) \supseteq \{a_=\}$. We define $\mu^= \coloneqq \frac{\mu^{-=} + \mu^{=+}}{2}$. It holds $\BR(\mu^=) = \{ a_=\}$. Hence,
        \begin{align*}
            \min_{p_L \in \upop}  U_F(p_L, a_=) &\overset{(i)}{=} U_F(\mu^=, a_=)\\
            &\overset{(ii)}{=} \max_{p_F \in \Delta(A_F)}  U_F(\mu^=, p_F)\\
            &\ge \min_{p_L \in \upop} \max_{p_F \in \Delta(A_F)}  U_F(p_L, p_F)\\
            &\overset{(iii)}{=} \max_{p_F \in \Delta(A_F)}  \min_{p_L \in \upop} U_F(p_L, p_F).
        \end{align*}
        \begin{enumerate}[(i)]
            \item The follower's action $p_F = a_=$ ties all leader's actions $p \in \upop$ by definition.
            \item Because $\max_{p_F \in \Delta(A_F)}  U_F(\mu^=, p_F) =  U_F(\mu^=, p^*_F)$ for any $p^*_F \in \BR(q)$ and $a_= \in \BR(q)$.
            \item Von Neumann's minmax theorem.
        \end{enumerate}
        Now let us argue that there is no other action $q \in \Delta(A_F)$ such that $q \in \BR^a(\upop)$: since we assumed that there are no weakly dominated strategies for follower,
        \begin{align*}
            U(\mu^=, a_=) > U(\mu^=, a), 
        \end{align*}
        for all $a \in A_F \setminus \{ a_=\}$.
    \end{enumerate}
\end{proof}

Lemma~\ref{lemma:N2 is exhaustive} shows that the ambiguous commitment sets $[\ell,u]$ with $(\ell, u) \in N^2$, where $N$ is defined as in Algorithm~\ref{alg:optimal ambiguous Stackelberg strategy}, exhaustively induce all possible response patterns.
\begin{lemma}
\label{lemma:N2 is exhaustive}
    Let $\cG$ be a coupled $2 \times m$-Stackelberg game. For any two points $\ell \le u$ in $[0,1]$ there exist $\underline{p}, \overline{p}' \in N$, where $N$ is defined as in Algorithm~\ref{alg:optimal ambiguous Stackelberg strategy}, such that for all $F \in \cF$
    \begin{align*}
        \BR^a_F([\ell, u]) = \BR^a_F([\underline{p}, \overline{p}]).
    \end{align*}
\end{lemma}
\begin{proof}
    If $\ell \in M$ (respectively $u \in M$), then $\underline{p} \coloneqq \ell$ (respectively $\overline{p} \coloneqq u$). Otherwise, we define $\underline{p}$ as the point in $M_\epsilon$ which minimizes the distance to $\ell$ and $\overline{p}$ as the point in $M_\epsilon$ which minimizes the distance to $u$. In case, there are two points in $M_\epsilon$ equally close, we choose arbitrarily.

    It is now a consequence of the construction of $M_\epsilon$ that none of the endpoints is pushed across a point $\mu^F_i$ in the response landscape. Formally, for any $F \in \cF$ and any $i \in \{ 1, \ldots, n-1\}$, (a) if $\ell < \mu^F_{i}$, then $\underline{p} < \mu^F_{i}$, and (b) if $\ell > \mu^F_{i}$, then $\underline{p} > \mu^F_{i}$. Note that we ruled out the case that $\ell = \mu^F_{i}$ for some $F \in \cF$ and $i \in \{ 1, \ldots , n-1\}$. The argument is is straight forward: let $\underline{p} \le \ell$, then (a) is clear. For case (b), note that for every $\ell > \mu^F_i$, $\ell$ is closer to $\mu^F_i + \epsilon$ than to $\mu^F_i - \epsilon$. An analogous argument holds for the second statement. Furthermore, it particularly holds for $u$ and $\overline{p}$ as well. Hence, we can apply Lemma~\ref{lemma:maxmin best responses -- empty equal action} or Lemma~\ref{lemma:maxmin best responses -- non-empty equal action}, which show that for each follower $F \in \cF$,
    \begin{align*}
        \BR^a_F([\ell, u]) = \BR^a_F([\underline{p}, \overline{p}]).
    \end{align*}
\end{proof}

\begin{lemma}
    Let $\cG$ be a coupled $2\times m$-Stackelberg game. Let $N$ be defined as in Algorithm~\ref{alg:optimal ambiguous Stackelberg strategy}. Let $C \coloneqq \max_{F \in \cF} \max_{a|F \in A_F} |U_{L_F}(ii, a_F) - U_{L_F}(i, a_F)|$, and $\epsilon > 0$ small enough. For any interval $[v,w] \subseteq [0,1]$ there exists an interval $[\ell, u]$ with $\ell, u \in N$ such that
    \begin{align*}
        \min_{p_L \in [\ell, u] } \left\langle p_L, \sum_{F \in \mathcal{F}} U_{L_F}(\cdot, b_F) \right\rangle \ge \min_{p_L \in [v,w] } \left\langle p_L, \sum_{F \in \mathcal{F}} U_{L_F}(\cdot, b_F) \right\rangle - \epsilon |\cF| C.
    \end{align*}
    where $b_F \coloneqq s(\BR^a_F([v,w]))$ for all $F \in \cF$, and $BR^a_F([v,w]) = BR^a_F([\ell,u])$. Without loss of generality we assume that
    \begin{align*}
        \left\langle p_L, \sum_{F \in \mathcal{F}} U_{L_F}(\cdot, b_F) \right\rangle,
    \end{align*}
    is linearly increasing. The argument analogously runs through in the decreasing case.
\end{lemma}
\begin{proof}
\label{proof:polytime algorithm gives optimal commitment}
    \begin{description}
        \item[Case] $v,w \notin M$
        \begin{enumerate}[(a)]
            \item Let us assume that there is no $\mu^F_i \in M$ such that $v < \mu^F_i < w$. We define the next point in $M$ right of $w$ as $\mu^r$. The next point in $M$ left of $v$ as $\mu^l$.

            By choice of $\epsilon$ we know that $\BR^a_F([v,w]) = \BR(\mu^l + \epsilon) = \BR_F(\mu^r - \epsilon)$ for all $F \in \cF$ (Lemma~\ref{lemma:maxmin best responses -- empty equal action} case (a) and (f), Lemma~\ref{lemma:maxmin best responses -- non-empty equal action} case (a) and (b), and Lemma~\ref{lemma:best response landscape})\footnote{We refer here to the cases in the lemmas which potentially apply. Not all cases have to apply simultaneously. For the sake of simplicity, we keep this notation throughout the following arguments.}. In particular, $\mu^l + \epsilon, \mu^r - \epsilon \in N$. Let $b_F \coloneqq s(\BR^a_F([v,w]))$. Hence,
            \begin{align*}
                \hat{W}(s) &\ge \max_{p_L \in \{ \mu^l + \epsilon, \mu^r - \epsilon\} } \left\langle p_L, \sum_{F \in \mathcal{F}} U_{L_F}(\cdot, b_F) \right\rangle\\
                &\ge \left\langle \mu^r - \epsilon, \sum_{F \in \mathcal{F}} U_{L_F}(\cdot, b_F) \right\rangle\\
                &= (\mu^r - \epsilon) \sum_{F \in \mathcal{F}} U_{L_F}(ii, b_F) + (1-(\mu^r - \epsilon)) \sum_{F \in \mathcal{F}} U_{L_F}(i, b_F)\\
                &= \left\langle \mu^r, \sum_{F \in \mathcal{F}} U_{L_F}(\cdot, b_F) \right\rangle - \epsilon \left(\sum_{F \in \mathcal{F}} U_{L_F}(ii, b_F) - \sum_{F \in \mathcal{F}} U_{L_F}(i, b_F)\right)\\
                &= \left\langle \mu^r, \sum_{F \in \mathcal{F}} U_{L_F}(\cdot, b_F) \right\rangle - \epsilon |\cF| C\\
                &\ge \min_{p_L \in [v,w]} \left\langle p_L, \sum_{F \in \mathcal{F}} U_{L_F}(\cdot, b_F) \right\rangle - \epsilon |\cF| C.
            \end{align*}
            In case the linear mapping is decreasing, an analogous argument provides the same statement.
            
            \item Let us assume that there is a single $\mu^F_i \in M$ such that $v < m^F_i < w$. By choice of $\epsilon$ it holds $\BR^a([v,w]) = \BR^a([\mu^F_i - \epsilon, \mu^F_i + \epsilon])$ (Lemma~\ref{lemma:maxmin best responses -- empty equal action} case (a),(e) and (f), Lemma~\ref{lemma:maxmin best responses -- non-empty equal action} case (a) and (b), and Lemma~\ref{lemma:best response landscape}). In particular, $\mu^F_i - \epsilon, \mu^F_i + \epsilon \in N$. For $b_F \coloneqq s(\BR^a_F([v,w]))$ this gives,
            \begin{align*}
                \hat{W}(s) &\ge \min_{p_L \in [\mu^F_i - \epsilon, \mu^F_i + \epsilon] } \left\langle p_L, \sum_{F \in \mathcal{F}} U_{L_F}(\cdot, b_F) \right\rangle\\
                &= \left\langle \mu^F_i - \epsilon, \sum_{F \in \mathcal{F}} U_{L_F}(\cdot, b_F) \right\rangle\\
                &= \left\langle \mu^F_i, \sum_{F \in \mathcal{F}} U_{L_F}(\cdot, b_F) \right\rangle - \epsilon |\cF| C\\
                &\ge \min_{p_L \in [v,w]} \left\langle p_L, \sum_{F \in \mathcal{F}} U_{L_F}(\cdot, b_F) \right\rangle - \epsilon |\cF| C.
            \end{align*}
            
            \item Let us assume that there are $\mu^F_i, \mu^{G}_{j} \in M$ such that $v < m^F_i < \mu^{G}_{j} < w$. By choice of $\epsilon$ it holds $\BR^a([v,w]) = \BR^a([\mu^F_i - \epsilon, \mu^G_j + \epsilon])$ (Lemma~\ref{lemma:maxmin best responses -- empty equal action} case (a),(e) and (f), Lemma~\ref{lemma:maxmin best responses -- non-empty equal action} case (a),(b) and (c), and Lemma~\ref{lemma:best response landscape}). In particular, $\mu^F_i - \epsilon, \mu^G_j + \epsilon \in N$.  For $b_F \coloneqq s(\BR^a_F([v,w]))$ this gives,
            \begin{align*}
                \hat{W}(s) &\ge \min_{p_L \in [\mu^F_i - \epsilon, \mu^G_j + \epsilon] } \left\langle p_L, \sum_{F \in \mathcal{F}} U_{L_F}(\cdot, b_F) \right\rangle\\
                &= \left\langle \mu^F_i - \epsilon, \sum_{F \in \mathcal{F}} U_{L_F}(\cdot, b_F) \right\rangle\\
                &= \left\langle \mu^F_i, \sum_{F \in \mathcal{F}} U_{L_F}(\cdot, b_F) \right\rangle - \epsilon |\cF| C\\
                &\ge \min_{p_L \in [v,w]} \left\langle p_L, \sum_{F \in \mathcal{F}} U_{L_F}(\cdot, b_F) \right\rangle - \epsilon |\cF| C.
            \end{align*}
        \end{enumerate}
        \item[Case] $v \in M, w \notin M$
        \begin{enumerate}[(a)]
            \item Let us assume that there is no $\mu^F_i \in M$ such that $v < \mu^F_i  < w$. By choice of $\epsilon$ it holds $\BR^a([v,w]) = \BR^a([v, v + \epsilon])$ (Lemma~\ref{lemma:maxmin best responses -- empty equal action} case (a),(b),(d) and (f), Lemma~\ref{lemma:maxmin best responses -- non-empty equal action} case (a) and (b), and Lemma~\ref{lemma:best response landscape}). In particular, $v, v+ \epsilon \in N$. For $b_F \coloneqq s(\BR^a_F([v,w]))$ this gives,
            \begin{align*}
                \hat{W}(s)  &\ge \min_{p_L \in [v, v + \epsilon] } \left\langle p_L, \sum_{F \in \mathcal{F}} U_{L_F}(\cdot, b_F) \right\rangle\\
                &= \left\langle v, \sum_{F \in \mathcal{F}} U_{L_F}(\cdot, b_F) \right\rangle\\
                &\ge \min_{p_L \in [v,w]} \left\langle p_L, \sum_{F \in \mathcal{F}} U_{L_F}(\cdot, b_F) \right\rangle.
            \end{align*}

            \item Let us assume that there is a $\mu^F_i \in M$ such that $v < \mu^F_i  < w$. Let $\mu^r \in M$ be the right most element of $M$ such that $\mu^r < w$. By choice of $\epsilon$ it holds $\BR^a([v,w]) = \BR^a([v, \mu^r + \epsilon])$ (Lemma~\ref{lemma:maxmin best responses -- empty equal action} case (a),(b),(d),(e) and (f), Lemma~\ref{lemma:maxmin best responses -- non-empty equal action} case (a) and (b), and Lemma~\ref{lemma:best response landscape}). In particular, $v, \mu^r + \epsilon \in N$.  For $b_F \coloneqq s(\BR^a_F([v,w]))$ this gives,
            \begin{align*}
                \hat{W}(s)  &\ge \min_{p_L \in [v, \mu^r + \epsilon] } \left\langle p_L, \sum_{F \in \mathcal{F}} U_{L_F}(\cdot, b_F) \right\rangle\\
                &= \left\langle v, \sum_{F \in \mathcal{F}} U_{L_F}(\cdot, b_F) \right\rangle\\
                &\ge \min_{p_L \in [v,w]} \left\langle p_L, \sum_{F \in \mathcal{F}} U_{L_F}(\cdot, b_F) \right\rangle.
            \end{align*}
        \end{enumerate}
        \item[Case] $v \notin M, w \in M$
        \begin{enumerate}[(a)]
            \item Let us assume that there is no $\mu^F_i \in M$ such that $v < \mu^F_i  < w$. By choice of $\epsilon$ it holds $\BR^a([v,w]) = \BR^a([w - \epsilon, w])$ (Lemma~\ref{lemma:maxmin best responses -- empty equal action} case (a),(b),(d) and (f), Lemma~\ref{lemma:maxmin best responses -- non-empty equal action} case (a) and (b), and Lemma~\ref{lemma:best response landscape}). In particular, $w - \epsilon, w \in N$. For $b_F \coloneqq s(\BR^a_F([v,w]))$ this gives,
            \begin{align*}
                \hat{W}(s)  &\ge \min_{p_L \in [w - \epsilon, w] } \left\langle p_L, \sum_{F \in \mathcal{F}} U_{L_F}(\cdot, b_F) \right\rangle\\
                &= \left\langle w - \epsilon, \sum_{F \in \mathcal{F}} U_{L_F}(\cdot, b_F) \right\rangle\\
                &= \left\langle w, \sum_{F \in \mathcal{F}} U_{L_F}(\cdot, b_F) \right\rangle - \epsilon |\cF| C\\
                &\ge \min_{p_L \in [v,w]} \left\langle p_L, \sum_{F \in \mathcal{F}} U_{L_F}(\cdot, b_F) \right\rangle  - \epsilon |\cF| C.
            \end{align*}

            \item Let us assume that there is a $\mu^F_i \in M$ such that $v < \mu^F_i  < w$. Let $\mu^l \in M$ be the left most element of $M$ such that $v < \mu^l$. By choice of $\epsilon$ it holds $\BR^a([v,w]) = \BR^a([\mu^l + \epsilon, w]) = (b_F)_{F \in \cF}$  (Lemma~\ref{lemma:maxmin best responses -- empty equal action} case (a),(b),(d),(e) and (f), Lemma~\ref{lemma:maxmin best responses -- non-empty equal action} case (a) and (b), and Lemma~\ref{lemma:best response landscape}). In particular, $\mu^l - \epsilon, w \in N$.  For $b_F \coloneqq s(\BR^a_F([v,w]))$ this gives,
            \begin{align*}
                \hat{W}(s)  &\ge \min_{p_L \in [\mu^l - \epsilon, w] } \left\langle p_L, \sum_{F \in \mathcal{F}} U_{L_F}(\cdot, b_F) \right\rangle\\
                &= \left\langle \mu^l - \epsilon, \sum_{F \in \mathcal{F}} U_{L_F}(\cdot, b_F) \right\rangle\\
                &= \left\langle \mu^l, \sum_{F \in \mathcal{F}} U_{L_F}(\cdot, b_F) \right\rangle - \epsilon |\cF| C\\
                &\ge \min_{p_L \in [v,w]} \left\langle p_L, \sum_{F \in \mathcal{F}} U_{L_F}(\cdot, b_F) \right\rangle - \epsilon |\cF| C.
            \end{align*}
        \end{enumerate}
        \item[Case] $v,w \in M$
        \begin{align*}
            \hat{W}(s) &\ge \min_{p_L \in [v,w]} \left\langle p_L, \sum_{F \in \mathcal{F}} U_{L_F}(\cdot, b_F) \right\rangle.
        \end{align*}
    \end{description}
\end{proof}

\subsection{Additional Proofs from Section~\ref{sec:hardness}}

\begin{theorem}
For all tie-breaking rules, there exists a coupled Stackelberg game $\cG$ such that the problem of determining the optimal pure ambiguous commitment is NP-hard. 
\label{thm:pure_hard}
\end{theorem}
\begin{proof}
\emph{Note: throughout this proof we utilize the fact that when multiple leader actions have the exact same follower utility vectors, the maxmin distribution of the follower can be computed by considering all equivalent leader actions as a single action.}

Let us assume we have a black-box algorithm $\A$ that takes as input a coupled Stackelberg game and outputs the optimal pure ambiguous commitment. We will use this to solve min-vertex cover. Given a graph $G$ with $n_{v}$ vertices, we will solve min-vertex cover for each connected  component separately. Thus, we can assume that the input to $G$ is connected. The input to $\A$ is as follows:
\begin{description}

\item[Cover Games] $k = n_{v}$ games of size $n_{v}$ x $2$, each associated with a particular vertex. The payoff for the leader is the same in all games: regardless of the leader's action, if the follower plays $a$, the leader gets payoff $1$, and if the follower plays $b$, the leader gets payoff $0$. From the follower in game $C_i$'s perspective, if vertex $j$ is adjacent to vertex $i$ or $j = i$, $u_{i}(j,a) = 0$ and $u_{i}(j,b) = -1$. Otherwise, $u_{i}(j,a) = 0$ and $u_{i}(j,b) = 1$. 
 \begin{table}[h]
        \centering
        \begin{tabular}{c|c c}
            $C_i$ & $a$ & $b$\\ \hline
            $a_L = i$ & 0 &  -1\\
            $a_L \in adj(i)$ & 0 &  -1\\
            $a_L \neq i, a_{L \notin adj(i)}$ & 0 & 1
        \end{tabular}
  \end{table} These games will represent the constraint that, given that the output must represent a vertex cover. 
\item[Minimization Game] One game of size $n_{v}$ x $n_{v}$. For both the leader and the follower, their payoffs in $M$ are the identity matrix. This game encodes the constraint that the output must represent a \emph{min} vertex cover. 
\end{description}

We will prove that the output of $\A$ on this input, when re-interpreted as a set of vertices, is the min vertex cover. 

\begin{lemma}
The maxmin strategies of the followers in the Cover games are always pure strategies, and the output of $\cA$ involves all the agents in these games playing action $a$ iff the strategies in $S$ correspond to a vertex cover.
\end{lemma}
\begin{proof}
To see this, note that if some vertex $i$ is not covered, the utility matrix for the follower in game $C_i$ includes only actions that strictly incentivize action $b$. Therefore, he will play pure strategy $b$. On the other hand, if the vertex is covered, the follower's utility matrix looks as follows:

 \begin{table}[h]
        \centering
        \begin{tabular}{c|c c}
             & $a$ & $b$\\ \hline
            $a_L = i$ & 0 &  -1\\
            $a_L \in adj(i)$ & 0 &  -1\\
            $a_L \neq i, a_{L \notin adj(i)}$ & 0 & 1
        \end{tabular}
  \end{table}

Given some distribution $p(a), p(b)$ of the follower, his payoff against a leader playing action $i$ is $-p(b)$, while his payoff against a leader playing any other action is $p(b)$. Therefore the worst-case for any follower distribution is attained against action $i$, so the follower's maxmin is the best response to action $i$. This is the pure strategy $a$. Therefore, each follower will either play $a$ or $b$, and he will play $a$ iff his corresponding vertex is covered in the pure ambiguous set. 

Next, we will show that the output to $\A$ must be a vertex cover. Assume for contradiction that this is not the case. Then we have some optimal pure ambiguous set $S$ which, from above, has at least one follower $i$ play action $b$. The payoff of the leader is equal to the number of followers that play action $a$, plus $\frac{1}{|S|}$ (Lemma~\ref{lemma: minimization game payoff}). Therefore, the payoff of $S$ is at most $ (n_{v} - 1) + 1 = n_{v}$. Now, consider the payoff of full ambiguity. Every follower is covered, so the payoff is $n_{v} + \frac{1}{n_{v}}$. This is larger, and therefore $S$ cannot be optimal. This proves a contradiction, and therefore if the output to $\A$ is integral, it must be a vertex cover.
\end{proof}

%\begin{lemma}The pure strategies output by $\A$ must correspond to a vertex cover in $G$, where $S_{i}$ corresponds to $V_{i}$.
%\end{lemma}

\begin{lemma}The pure strategies output by $\A$ must correspond to a min-vertex cover in $G$.
\end{lemma}
\begin{proof}
Note that the utility for the leader for every vertex cover is $n_{v} + \frac{1}{|S|}$ (Lemma~\ref{lemma: minimization game payoff}). This is maximized when $|S|$ is minimized. Therefore, the solution output by $\A$ must be not only a vertex cover but a min-vertex cover. 
\end{proof}
\end{proof}

\subsection{On the Structure of Optimal Commitments}
\label{app:structure}
Many of the examples we discuss showing the power of ambiguous commitments are utilizing \emph{full} ambiguity; in other words, the leader's ambiguous set is all her pure strategies. If it was the case that this is always the best form of ambiguity to use, then an algorithm for the best ambiguous commitment would be trivial. However, in this section we show that there are settings where the optimal ambiguous commitment is neither a singleton nor the entirety of the pure strategy set. It can be a subset of the pure strategies, or even have extreme points that are not pure strategies. 

\begin{theorem}
There is a general-sum game where the gap between the best ambiguous commitment and the best of (full ambiguity, decoupled stackelberg) is unbounded. \label{thm:pure_unbounded}
\end{theorem}

\begin{proof}
 Table~\ref{tab:piib-F1}, Table~\ref{tab:piib-F2}, Table~\ref{tab:piib-F1-2} and Table~\ref{tab:piib-F2-2} define the general-sum game. The decoupled Stackelberg value is $\frac{2c}{B}$. The value of full ambiguity is $0$. But by committing to $\{1,2\}$, the leader gets utility $c$. By taking $B$ sufficiently large, this gap is unbounded. 
\begin{table}[h]
    \centering
    \begin{minipage}{0.45\textwidth}
        \centering
        \begin{tabular}{c|c c}
             $F_1$ & $b_1$ & $b_2$\\ \hline
            $a_L = 0$ & B &  1\\
            $a_L = 1$ & 0 & 1 \\
            $a_L = 2$ & 0 & -1 \\
        \end{tabular}
        \caption{Follower $F_1$'s payoffs.}
        \label{tab:piib-F1}
    \end{minipage}%
    \hfill
    \begin{minipage}{0.45\textwidth}
        \centering
        \begin{tabular}{c|c c}
             $F_2$ & $c_1$ & $c_2$ \\ \hline
            $a_L = 0$ & 1 & 0\\
            $a_L = 1$ & 1 & B\\
            $a_L = 2$ & -1 & 0
        \end{tabular}
        \caption{Follower $F_2$'s payoffs.}
        \label{tab:piib-F2}
    \end{minipage}
    \vspace{0.5cm} % Adds vertical space between rows of tables
    \begin{minipage}{0.45\textwidth}
        \centering
        \begin{tabular}{c|c c}
             $F_1$ & $b_1$ & $b_2$\\ \hline
            $a_L = 0$ & 0 &  c\\
            $a_L = 1$ & 0 & 0 \\
            $a_L = 2$ & 0 & 0 \\
        \end{tabular}
        \caption{Leader's payoff against $F_1$.}
        \label{tab:piib-F1-2}
    \end{minipage}%
    \hfill
    \begin{minipage}{0.45\textwidth}
        \centering
        \begin{tabular}{c|c c}
             $F_1$ & $b_1$ & $b_2$\\ \hline
            $a_L = 0$ & 0 &  0\\
            $a_L = 1$ & c & 0 \\
            $a_L = 2$ & 0 & 0 \\
        \end{tabular}
        \caption{Leader's payoff against $F_2$.}
        \label{tab:piib-F2-2}
    \end{minipage}
\end{table}
 
\end{proof}

\begin{theorem}
There is a $2 \times m$-game in which the optimal coupled commitment is fractional. \label{thm:frac_opt}
\end{theorem}

\begin{proof}
Table~\ref{tab:piib-F1-2x m}, Table~\ref{tab:piib-F2-2x m}, Table~\ref{tab:piib-F1-2-2x m} and Table~\ref{tab:piib-F2-2-2x m} define the $2\times m$-game. We will consider three cases:
\begin{itemize}
\item The leader commits to a single distribution for both followers. If this distribution incentivizes $F_{1}$ to play $b_{2}$, then

\begin{align*}
& -\frac{2}{3}Pr(a_{L} = 0) + Pr(a_{L} = 1) < 0 \\
& \implies  Pr(a_{L} = 0) > \frac{3}{5} 
\end{align*}

But then the utility of $F_{2}$ for action $c_1$ is at least $\frac{3}{5} - \frac{2}{3}(\frac{2}{5}) = \frac{1}{3} $, which is strictly greater than his utility for $c_{2}$. Therefore, for any distribution that $F_1$ responds to with $b_{2}$, $F_{2}$ will respond with $c_{1}$. Therefore the maximum value of the coupled game is $3$. 
\item The leader commits to full ambiguity. Then both followers will play their second action. The payoff for the leader is now computed in the worst case over her commitment, which is $4$. 
\item The leader commits to the interval $[0.4 - \epsilon,0.6 + \epsilon]$. Then both the followers will still respond with their second action. However, now the worst-case payoff for the leader is 
\begin{align*}
& \min((0.4 - \epsilon)5 + (0.6 + \epsilon)4, (0.6 + \epsilon)5 + (0.4 - \epsilon)4) \\
& = \min(4.4 - \epsilon,4.6 + \epsilon) = 4.4 - \epsilon
\end{align*}

This is strictly higher than any non-fractional coupled commitment. 
\end{itemize}

\begin{table}[h]
    \centering
    \begin{minipage}{0.45\textwidth}
        \centering
        \begin{tabular}{c|c c c}
             $F_1$ & $b_1$ & $b_2$ & $b_3$\\ \hline
            $a_L = 0$ & 1 &  0 & $-\frac{2}{3}$\\
            $a_L = 1$ & -4 & 0 & 1 \\
        \end{tabular}
        \caption{Follower $F_1$'s payoffs.}
        \label{tab:piib-F1-2x m}
    \end{minipage}%
    \hfill
    \begin{minipage}{0.45\textwidth}
        \centering
        \begin{tabular}{c|c c c}
             $F_2$ & $c_1$ & $c_2$ & $c_3$\\ \hline
            $a_L = 0$ & 1 & 0 & -4\\
            $a_L = 1$ & $-\frac{2}{3}$ & 0 & 1\\
        \end{tabular}
        \caption{Follower $F_2$'s payoffs.}
        \label{tab:piib-F2-2x m}
    \end{minipage}
    \vspace{0.5cm} % Adds vertical space between rows of tables
    \begin{minipage}{0.45\textwidth}
        \centering
        \begin{tabular}{c|c c c}
             $F_1$ & $b_1$ & $b_2$ & $b_3$\\ \hline
            $a_L = 0$ & 0 &  2 & 0\\
            $a_L = 1$ & 0 & 2 & 0 \\
        \end{tabular}
        \caption{Leader's payoff against $F_1$.}
        \label{tab:piib-F1-2-2x m}
    \end{minipage}%
    \hfill
    \begin{minipage}{0.45\textwidth}
        \centering
        \begin{tabular}{c|c c c}
             $F_1$ & $c_1$ & $c_2$ & $c_3$\\ \hline
            $a_L = 0$ & 0 &  3 & 0\\
            $a_L = 1$ & 0 & 2 & 0 \\
        \end{tabular}
        \caption{Leader's payoff against $F_2$.}
        \label{tab:piib-F2-2-2x m}
    \end{minipage}
\end{table}
   
\end{proof}

\end{document}